\documentclass[preprint,secnumarabic,amsmath,amssymb]{revtex4}
\usepackage{amsmath}

\newtheorem{remark}{Remark}[section]
\usepackage{graphicx}
\usepackage{epsf}
\usepackage{bm}
\usepackage{amsmath}
\usepackage{amsfonts}
\usepackage{amssymb}
\usepackage{epstopdf}
\usepackage{natbib}
\usepackage{color}

\begin{document}
\tolerance=5000
\noindent

\title{Reheating in quintessential inflation via gravitational production of heavy massive particles: A detailed analysis}

\author{Jaume Haro$^{1}$\footnote{E-mail: jaime.haro@upc.edu}, Weiqiang Yang$^{2}$\footnote{E-mail: d11102004@163.com} and Supriya Pan$^{3}$\footnote{E-mail: supriya.maths@presiuniv.ac.in}}

\affiliation{$^{2}$Departament de Matem\`atiques, Universitat Polit\`ecnica de Catalunya, Diagonal 647, 08028 Barcelona, Spain}

\affiliation{$^{2}$Department of Physics, Liaoning Normal University, Dalian, 116029, P. R. China.}

\affiliation{$^{3}$Department of Mathematics, Presidency University, 86/1 College Street, Kolkata 700073, India.}


\begin{abstract}
An improved version of the well-known Peebles-Vilenkin model unifying early inflationary era to current cosmic acceleration, is introduced in order 
to match with the theoretical values of the spectral quantities provided by it with the  recent observational data about the early universe.  Since the model presents a sudden phase transition, we consider  the simplest way to reheat the universe $-$ via the gravitational production of heavy massive particles $-$ which assuming that inflation starts at GUT scales $\sim 10^{16}$ GeV,  allows us 
to use the Wentzel-Kramers-Brillouin (WKB) approximation and consequently this enables us to perform all the calculations in an analytic way. Our results show that the model leads to a maximum temperature at the TeV regime, and passes the bounds to ensure the success of the Big Bang Nucleosynthesis. Finally, we have constrained the quintessence piece of the proposed improved version of the Peebles-Vilenkin model using various astronomical datasets available at present.  
\end{abstract}

\vspace{0.5cm}



\maketitle

\section{ Introduction}

The inflationary paradigm, an early accelerating phase of the universe, was implemented into the Big Bang Cosmology (BBC) with an aim to answer several deficiencies associated to BBC, such as the horizon problem, flatness or the primordial monopole problem as well as some more \cite{guth, linde}. Soon after the introduction of the inflationary paradigm, this mechanism was used to explain the primordial cosmological perturbations \cite{chibisov, starobinsky, pi, bardeen}  with great agreement with the recent observational data from Planck \cite{Planck}. Certainly, the agreement with the observational data is a clear manifestation of the success of the inflationary paradigm. According to the earlier studies \cite{guth, linde,chibisov, starobinsky, pi, bardeen} (also see \cite{Linde:1982uu,Burd:1988ss,Barrow:1990vx,Barrow:1994nt}), the simplest viable scenario that depicts our universe comes through the inflation. While on the other hand, several attempts were also made aiming to obtain viable alternative cosmologies such as the non-singular bouncing cosmologies \cite{novello, lehners, peter, brandenberger, cai, nojiri}, where the Big Bang singularity, which is an unsolved problem of the inflationary cosmology, is replaced by a non-singular Big Bounce. In the non-singular bouncing cosmologies, the basic methodology is to obtain a dynamical universe that evolves from a contracting  phase of the universe to its expanding one \cite{wilson, cai3, Haro13, Haro13a, wilson2, cai1, Haro16, haa17} (also see \cite{deHaro:2018cpl,Das:2018bzx}) and thus, it naturally avoids the Big Bang singularity.

What is important to point out is that a viable inflationary scenario needs a reheating mechanism in order to match with the Hot Big Bang (HBB) cosmology, because all the 
pre-existing particles are diluted after the end of inflation due to the large size increase of our universe patch. This is not a trivial point, and in the case of standard inflation, i.e., when the potential has a well deep, after inflation, the inflaton field oscillates. As a consequence, the inflaton field releases its energy to produce the massive particles that after its decay and thermalization, reheat the universe. This mechanism has been widely studied in a series of works by several investigators \cite{kls, kls1, gkls, stb, Basset}. However, after the discovery of the current cosmic acceleration \cite{riess, perlmutter}, a class of cosmological models attempting to unify the early- and late- accelerating expansions, 
the so-called  {\it quintessential inflation} models \cite{Spokoiny, pr, pv} appeared where by constructions, the models are able to produce an early- and a late- accelerated expansions of the universe (also see \cite{deHaro:2016hpl,hap,deHaro:2016hsh,deHaro:2016ftq,deHaro:2017nui,Geng:2017mic,AresteSalo:2017lkv, Haro:2015ljc, hossain1,hossain2,hossain3,hossain4}). 
A natural behaviour of the {\it quintessential inflation} models is, 
the potential of the inflaton field does not have a local minima and thus, the inflaton field does not oscillate. For these models, to reheat  the  universe, a phase transition where the adiabatic regime is broken, 
is needed in order to create an enough amount of particles which after decays and interactions with other fields, form a thermal relativistic plasma whose energy density becomes dominant.  This mechanism of particle creation can be obtained in different ways, such as the  gravitational particle production \cite{Parker,gm,glm,gmm, ford, Birrell1, Zeldovich, dimopoulos0, hashiba}, instant preheating
\cite{fkl0, fkl, dimopoulos, vardayan},  curvaton reheating \cite{LU,FL,ABM}, 
production of massive particles where the masses of the massive particles 
depend on the inflaton field \cite{rubio}, or, the 
production of massive particles with a self-interaction and coupled to gravity \cite{tommi}.

Another important question is related to the bounds of the reheating temperature. 
A lower bound is obtained recalling  that the
radiation dominated era is prior to the Big Bang  Nucleosynthesis (BBN) epoch which occurs in the  $1$ MeV  regime \cite{gkr}. 
As a consequence,  the reheating temperature has to be greater
than $1$ MeV.  The upper bounds may depend on the theory we are dealing with, for instance, in many supergravity and superstring theories containing particles, such as the gravitino or a modulus field, with only gravitational interactions, and thus
the late time decay of these relics products may disturb  
the success of the standard BBN \cite{lindley}.
This problem can be successfully removed if the  reheating temperature is
of the order of $10^9$ GeV \cite{eln}. In the present study,  we will accept this usual bound restricting  the reheating temperature staying between $1$ MeV and $10^9$ GeV.
On the other hand, one has to take into account that a viable reheating mechanism has to deal with the affectation of the Gravitational Waves (GWs) in the BBN success by satisfying the observational bounds coming from the overproduction of the GWs \cite{pv}, or related to the logarithmic spectrum of its energy density \cite{maggiore}.

In the present work, we consider an improved version of a well known {\it quintessential inflation} model, namely, the Peebles-Vilenkin potential \cite{pv}, where the inflationary piece is now changed to quadratic instead of quartic as in the original version. We do this change because a quartic potential leading to a scalar spectral index, $n_s$, and the ratio of tensor to scalar perturbations, $r$, do not enter in the marginalized joint confidence contour in the $2$-dimensional plane at $2\sigma$ confidence-level (CL) \cite{Planck}, but since in the quintessential inflation model, the number of $e$-folds is larger than in standard inflation, a quadratic potential leads to  theoretical values of $(n_s, r)$ that enter at $2\sigma$ CL  (see \cite{hap} for a detailed discussion). We also assume a pre-heating due to the gravitational production of heavy massive particles, which will decay in lighter ones to form a thermal relativistic plasma, because one can use the  Wentzel-Kramers-Brillouin (WKB) method or  approximation to calculate the vacuum modes. And in the case of a sudden phase transition, it is possible to calculate the energy density of the produced particles in an analytic way. Consequently, one becomes able to perform the calculations of the relevant quantities in an analytical way, such as the decaying time or the reheating temperature. We note that both the quantities that means decaying time and the reheating temperature depend on whether the decay occurs before or after the time at which the energy density of the background is equal to that of the produced particles, i.e., at the end of kination \cite{Joyce}. 
Finally, we show  that our model overpasses the constraints given by the production of GWs and leads to  reheating temperatures compatible with the BBN success.

Our study is structured as follows.
In Section \ref{sec-II} we apply the  WKB method to cosmology, showing how to approximate, during the adiabatic regimes, the vacuum modes using the $n^{\rm th}$ order  WKB approximation, and discussing when one can use it in the early universe. Section \ref{sec-III} is devoted to the study of the reheating via particle production of massive particles considering the improved version of the Peebles-Vilenkin model \cite{pv} with a sudden phase transition from inflation to kination where using the WKB approximation, we calculate the energy density of the produced particles, and consequently, we obtain the reheating temperature in two different situations, namely when the decay of the particles is before the end of kination and when it is after it. In Section \ref{sec-BBN} we consider the bounds imposed in order that GWs do not disturb the BBN, showing that our model overpasses them. After that in Section \ref{sec-data+results} we constrain the quintessence piece of the model using the latest astronomical data from various sources. Finally, we close the present work in Section \ref{sec-conclu} presenting a brief summary of the entire results.  
Last but not least, let us mention that, throughout the entire calculations presented in the next sections, the units used are $\hbar=c=1$, and we denote  the  reduced Planck's mass  by $M_{pl}\equiv \frac{1}{\sqrt{8\pi G}}\cong 2.4\times 10^{18}$ GeV.

\section{The  WKB approximation in cosmology}
\label{sec-II}

We begin our analysis considering a massive quantum field $\chi$ which is conformally coupled to gravity with the following Lagrangian \cite{Birrell}:
\begin{eqnarray}
{\mathcal L}=\frac{1}{2}\left(\chi_{\mu}\chi^{\mu}-m_{\chi}^2\chi^2-\frac{1}{6}R\chi^2\right),
\end{eqnarray} 
where  $m_{\chi}$ is the bare mass  of the quantum field and $R$ is the scalar quantity known as the Ricci curvature. The corresponding Klein-Gordon (K-G) equation following the variation of $\chi$,  is given by
\begin{eqnarray}
\left(-\nabla_{\mu}\nabla^{\mu}+m_{\chi}^2+ \frac{1}{6}R\right)\chi=0.
\end{eqnarray}

Now, in the background of a spatially flat Friedmann-Lema{\^\i}tre-Robertson-Walker (FLRW) spacetime, and working in the Fourier space, the K-G equation can be recast into 
\begin{eqnarray}\label{kg0}
\chi_k''+2{\mathcal H}\chi_k'+\left(k^2+m_{\chi}^2a^2+\frac{a''}{a}    \right)\chi_k=0,
\end{eqnarray}
where the prime denotes the derivative with respect the conformal time $\tau$,  and ${\mathcal H} \equiv a'/a$, is the conformal Hubble parameter.

In order to understand the equation (\ref{kg0}) clearly, it is useful to perform the following change of variable $\bar{\chi}_k=a\chi_k$, that gives 
\begin{eqnarray}\label{kg}
\bar{\chi}_k''+\left(k^2+m_{\chi}^2a^2   \right)\bar{\chi}_k=0,
\end{eqnarray}
which is the equation of an harmonic oscillator with 
time depend frequency $\omega_k(\tau)=\sqrt{k^2+m_{\chi}^2a^2(\tau)}$.
During the adiabatic regimes, 
 to calculate the $k$-vacuum mode, one can use the WKB approximation \cite{Haro}
\begin{eqnarray}
\bar{\chi}_{n,k}^{WKB}(\tau)\equiv
\frac{1}{\sqrt{2W_{n,k}(\tau)}}e^{-{i}\int^{\tau}W_{n,k}(\eta)d\eta},
\end{eqnarray}
where $n$ is the order of the approximation and $W_{n,k}(\tau)$ is calculated as follows (see for more details \cite{Winitzki}). 
First of all, instead of equation (\ref{kg}) we consider the following equation 
\begin{eqnarray}\label{kgepsilon}
\bar\epsilon\bar{\chi}_k''+\omega_k^2(\tau)\bar{\chi}_k=0,
\end{eqnarray}
where $\bar\epsilon$ is a dimensionless parameter that one may set  $\bar\epsilon=1$ at the end of calculations. 
Looking for a solution of (\ref{kgepsilon}) of the form 
\begin{eqnarray}
\bar{\chi}_{n,k}^{WKB}(\tau; \bar\epsilon)=
\frac{1}{\sqrt{2W_{n,k}(\tau;\bar\epsilon)}}e^{-\frac{i}{\bar\epsilon}\int^{\tau}W_{n,k}(\eta;\bar\epsilon)d\eta},
\end{eqnarray}
where $W_{0,k}(\tau;\bar\epsilon)\equiv \omega_k(\tau)$, 
after inserting it in (\ref{kgepsilon}) and  collecting the terms of order $\bar\epsilon^{2n}$  one gets the following iterative formula
\begin{eqnarray}
W_{n,k}(\tau;\bar\epsilon)= \mbox{ terms up to order } \bar\epsilon^{2n} \mbox{ of } 
\left( \sqrt{\omega_k^2(\tau)-\bar\epsilon^2 \left[\frac{1}{2} \frac{W''_{n-1,k}(\tau;\bar\epsilon)}{W_{n-1,k}(\tau;\bar\epsilon)} 
-\frac{3}{4}\frac{(W'_{n-1,k}(\tau;\bar\epsilon))^2}{W^2_{n-1,k}(\tau;\bar\epsilon)} \right]}         \right).
\end{eqnarray}

Then, as an example,  a simple calculation leads, after setting $\bar\epsilon=1$,  to 
\begin{eqnarray}\label{1approx}
W_{1,k}(\tau)=
\omega_k-\frac{1}{4}\frac{\omega''_{k}}{\omega^2_{k}}+\frac{3}{8}\frac{(\omega'_{k})^2}{\omega^3_{k}}.
\end{eqnarray}

On the other hand, 
the standard condition to guarantee the adiabatic regime is $\omega_k'\ll \omega_k^2$. For this condition one can approximate the modes by the zero order 
WKB approximation
\begin{eqnarray}
\bar{\chi}_{0,k}^{WKB}(\tau)\equiv
\frac{1}{\sqrt{2\omega_{k}(\tau)}}e^{-{i}\int^{\tau}\omega_{k}(\eta)d\eta},
\end{eqnarray}
but to use the $n^{\rm th}$-order approximation, one needs that the more general condition is,
\begin{eqnarray}
\left|\frac{d^n \omega_k}{d\tau^n}\right|\ll \omega_k^{n+1},
\end{eqnarray}
which is always satisfied when $H\ll m_{\chi}$.

The question that arises now is, when in the early universe,
one can apply the WKB approximation. It is well-known that at temperatures of the  order of the Planck's mass, quantum effects become very important  and the classical picture of the universe is not possible. However, at temperatures below $M_{pl}$, for example 
at GUT scales (i.e., when the temperature is of the order of $T\sim 4\times 10^{-3} M_{pl}\sim  10^{16}$ GeV), the beginning of the  Hot Big Bang (HBB) scenario is possible. Since for the  flat FLRW universe, the energy density of the universe, namely, $\rho$,  and the Hubble parameter $H$ are related through $\rho=3H^2M_{pl}^2$, and the temperature of the universe is related to the energy density via $\rho = (\pi^2/30)g_{*} T^4$, where  the degrees of freedom is,
$g_*=107, 90$ or $11$, for temperatures satisfying  respectively, $T\geq 175$ GeV, $175$ GeV $> T > 200$ MeV and $200$ MeV $\geq T \geq 1$ MeV (see for instance \cite{rg}); thus, one can conclude that a classical picture of the universe would be possible when $H\cong 5\times 10^{-5} M_{pl}\cong 10^{14}$ GeV. Then, 
if inflation starts at this scale, i.e., taking the value of the Hubble parameter at the beginning of inflation (denoted by $H_{beg}$) as $H_{beg}=5\times 10^{-5} M_{pl}$; assuming that the quantum $\chi$-field is in the vacuum at the beginning of inflation; 
and choosing  the mass of the  $\chi$-field one order greater than this value of  the Hubble parameter ($m_{\chi}=5\times 10^{-4} M_{pl}\cong 10^{15}$ GeV which is a mass of the same order as those of the vector mesons responsible to transform quarks into leptons in simple theories with  SU(5)  symmetry \cite{lindebook}),  one can apply the WKB approximation to calculate
 the renormalized  energy density of the vacuum, obtaining, after subtracting the adiabatic modes up to order four,  an energy density of the order   
 $H^6/m_{\chi}^2$ \cite{kaya} which is  subdominant compared to the energy density of the background $3H^2M_{pl}^2$.

\begin{remark}
The exact value of the energy density of the vacuum was calculated in \cite{bunch} assuming the case of an exact de Sitter phase in the flat FLRW space-time.
Choosing the vacuum  modes
\begin{eqnarray}\label{b178}
\bar{\chi}_{ k}(\tau)=C\sqrt{\frac{\pi\tau}{4}}H^{(2)}_{\nu}(k \tau),
\end{eqnarray}
with $\nu\equiv\sqrt{\frac{9}{4}-\frac{m_{\chi}^2}{H^2}-12\xi}$ and
$C\equiv e^{-i(\frac{\pi\nu}{2}+\frac{\pi}{4})}$, Bunch and Davies, using the point-splitting regularization obtained the energy density: 
\begin{eqnarray}\label{b179}
\rho_{\chi}=
&&\frac{1}{64\pi^2}\left\{m_{\chi}^2\left[
m_{\chi}^2+(12\xi-2)H^2\right]\left[\psi\left(\frac{3}{2}+\nu\right)+\psi\left(\frac{3}{2}-\nu\right)-\ln\left(\frac{m_{\chi}^2}{H^2}\right)\right]
\nonumber\right.\\&&\left.-m_{\chi}^2(12\xi-2)H^2-\frac{2}{3}m_{\chi}^2H^2-\frac{1}{2}(12\xi-2)^2H^4+\frac{H^4}{15}\right\},
\end{eqnarray}
where $\xi$ is the coupling constant with gravity and $\psi$ denotes the digamma function.  It is instructive to see that, when $m_{\chi}\gg H$,  the terms containing 
$ \ln\left(\frac{m_{\chi}^2}{H^2}\right) $, $m_{\chi}^2 H^2$ and $H^4$ cancel, 
and one obtains an energy density of the vacuum with
$\sim {\mathcal O}\left( \frac{H^6}{m_{\chi}^2}  \right)$.
\end{remark}

Finally, the evolution of the vacuum goes as follows: the $k$-vacuum mode  during the adiabatic regime could be approximated by $\bar{\chi}_{n,k}^{WKB}$, but when
the adiabatic regime breaks down during a period of time, the WKB approximation could not be used, and only at the end of this period, one could again use
it; but now the vacuum mode is a combination of positive and negative frequency modes which could be approximated by a linear combination of $\bar{\chi}_{n,k}^{WKB}$ and its conjugate of  the form $\alpha_{n,k}\bar{\chi}_{n,k}^{WKB} + \beta_{n,k}(\bar{\chi}_{n,k}^{WKB} )^*  $, where $\alpha$ and $\beta$ are the so-called {\it Bogoliubov coefficients}, and it is the manifestation  of the gravitational particle production. Basically, this  is  Parker's viewpoint of  particle creation in curved space-times  \cite{Parker}, where the $\beta$-Bogoliubov coefficient is the key ingredient to calculate  the energy density of the produced particles.

\section{Reheating in quintessence inflation via gravitational production of heavy massive particles}
\label{sec-III}

In order to deal with an analytically solvable problem, i.e., having an analytic expression of the $\beta$-Bogoliubov coefficient, we consider a sudden phase transition where the second derivative of the Hubble parameter is discontinuous, which happens for the following {\it improved version} of the well-known Peebles-Vilenkin  quintessential inflationary potential \cite{pv}
\begin{eqnarray}\label{PV}
V(\varphi)=\left\{\begin{array}{ccc}
\frac{1}{2}m^2\left(\varphi^2-M_{pl}^2+M^2\right)& \mbox{for}& \varphi\leq -M_{pl}\\
\frac{1}{2}m^2\frac{M^6}{(\varphi+M_{pl})^4+M^4}& \mbox{for}& \varphi\geq -M_{pl},
\end{array} \right.
\end{eqnarray}
where the free parameter $M$ can be approximated as $M\sim 20$ GeV (see \cite{pv} and \cite{haro18} for a detailed discussion on how the value of this parameter is obtained). 

Here, it is important to point out that
the inflationary piece of the original Peebles-Vilenkin potential is quartic, and thus the theoretical values of  spectral index and the ratio of tensor to scalar perturbations do not enter in the marginalized  joint confidence contour in the plane  $(n_s,r)$ at $2\sigma$ CL \cite{Planck}, without the presence of running \cite{hap}. 
This is the reason why we have changed the quartic part of the potential by the quadratic potential, whose spectral values, due to the fact that in the quintessential inflation, the number of $e$-folds for realistic models is between $63$ and $75$, do actually enter in this contour \cite{hap}. This is the main motivation behind this work.  

To calculate the mass of the inflaton, we use the theoretical and observational values of the power spectrum. The power spectrum of the curvature fluctuation in a co-moving coordinate system when the pivot scale leaves the Hubble radius  is given by \cite{btw}:
${\mathcal P}_{\zeta}\cong \frac{H_*^2}{8\pi^2 M_{pl}^2\epsilon_*}\sim 2\times 10^{-9}$ where $\epsilon=-\frac{\dot{H}}{H^2}\cong\frac{M_{pl}^2}{2}\left(\frac{V_{\varphi}}{V}\right)^2$, is the main slow roll parameter and the symbol ``$\ast$'' (pronounced as ``star'') means that the quantity is evaluated when the pivot scale leaves the Hubble radius, obtaining
\begin{eqnarray}
m^2\sim 3\times 10^{-9} \pi^2(1-n_s)^2   M_{pl}^2,
\end{eqnarray}
where we have used that for our model one has $\epsilon_*= \frac{2M_{pl}^2\varphi^2_*}{(\varphi^2_*-M_{pl}^2)^2}\cong
 \frac{2M_{pl}^2}{\varphi^2_*}$,  because $-\varphi_*\gg M_{pl}$ which means that, $\epsilon_*\cong \frac{1-n_s}{4}$, where $n_s$ denotes the spectral index,
 and during inflation one has, $H_*^2\cong \frac{1}{6M_{pl}^2}m^2\varphi_*^2$. 
Since from the recent observations by Planck \cite{Planck}, the value of the spectral index is constrained to be, $n_s=0.968\pm 0.006$ \cite{Planck}, thus, taking its central value  one gets, $m\cong 5\times 10^{-6} M_{pl}$, and as a consequence, if one assumes, as usual, that there is not any substantial drop of the energy density between the end of inflation and the beginning of kination,
and moreover, takes into account that, $\varphi_{end}=-\sqrt{2+\sqrt{3}}M_{pl}$,
then one finds that, $H_{kin}\sim H_{end}\cong \frac{m\sqrt{\varphi_{end}^2-M_{pl}^2}}{\sqrt{6}M_{pl}}=\sqrt{\frac{1+\sqrt{3}}{6}}{m} \cong 3\times10^{-6} M_{pl}$, where $H_{kin}$ and $H_{end}$ denote respectively the values of the Hubble parameter at the beginning of kination and at the end of inflation.

\begin{remark}
In the same way, one can obtain that the value of the Hubble parameter when the pivot scale leaves the Hubble radius is, $H_*\cong 3\times 10^{-5} M_{pl}$, which is, of course, between the values of the Hubble parameter at the beginning, $H_{beg}\cong 5\times 10^{-5} M_{pl}$, and at the end, $H_{end}\cong 3\times 10^{-6} M_{pl}$ of the inflation.

\end{remark}

On the other hand, one can easily calculate 
the effective Equation of State (EoS)  parameter which is equal to, $w_{eff}=-1-\frac{2\dot{H}}{3H^2}=-1+\frac{2}{3}\epsilon$, which means that 
for $\varphi\ll -M_{pl}$, one has $\epsilon\ll 1$ (slow-roll period) and then $w_{eff}\cong -1$. Immediately after  the end of inflation, which 
as we have already seen
occurs at  $\varphi_{end}=-\sqrt{2+\sqrt{3}}M_{pl}$, the universe suffers a phase transition from inflation to a kination regime \cite{Joyce}, which starts at $\varphi=-M_{pl}$, and where due to  the small value
of the parameter $M$,  the potential energy density is negligible compared to the kinetic one. This means that in the kination phase, 
 $w_{eff}\cong 1$. Note that at the end of the phase transition, i.e., at $\varphi=-M_{pl}$, the adiabatic regime is broken, because the second derivative of the Hubble parameter is discontinuous, and thus particles are produced. Effectively, 
 the derivative of the potential is discontinuous at $\varphi=-M_{pl}$, which means that due to the conservation equation, the second derivative of the 
inflaton field is discontinuous  at the beginning of kination. As a consequence, using the Raychaudhuri equation,
$\dot{H}=-\frac{\dot{\varphi}^2}{2M_{pl}^2}$, one can deduce that the second derivative of the Hubble parameter is also discontinuous at this time. 
Then, since during the kination regime the energy density of the scalar field decreases faster than the energy density of the produced particles,  eventually the energy density of the produced particles starts to dominate and the universe enters into the radiation phase which ends at the matter-radiation equality.
Finally, at the present time, due to the value of the parameter $M$, the energy density of the field, which is practically all potential, starts to dominate once again, and thus,  $\dot{H}\cong 0$, which means $w_{eff}\cong -1$, showing the current cosmic acceleration.

To perform all the calculations in an analytical way, in our case,  
we only need the first order WKB solution to approximate  the $k$-vacuum modes before and after the phase transition, and this is given by 
\begin{eqnarray}
\bar{\chi}_{1,k}^{WKB}(\tau)\equiv
\frac{1}{\sqrt{2W_{1,k}(\tau)}}e^{-{i}\int^{\tau}W_{1,k}(\eta)d\eta},
\end{eqnarray}
because $W_{1,k}$ (see eqn. (\ref{1approx}))  contains the first derivative of the Hubble parameter, and since the matching involves the derivative of the mode, the $\beta$-Bogoliubov coefficient does not vanish.

Before the  phase transition time, namely, $\tau_{kin}$, the  vacuum mode is depicted by $\chi_{1,k}^{WKB}(\tau)$, but after the phase transition this mode becomes a mixture of positive and negative frequencies of the form
$\alpha_k \chi_{1,k}^{WKB}(\tau)+\beta_k (\chi_{1,k}^{WKB})^*(\tau)$.
The $\beta_k$-Bogoliubov coefficient is obtained by matching both the expressions  and its derivatives at $\tau_{kin}$ \cite{haro18}
\begin{eqnarray}
 |\beta_k|^2\cong \frac{m^4_{\chi}m^6a_{kin}^{10}}{256(k^2+m_{\chi}^2a_{kin}^2)^5},
\end{eqnarray}
where we have introduced the notation $a_{kin}=a(\tau_{kin})$.

On the other hand, the vacuum average energy density of the $\chi$-field is given by \cite{Birrell}
 \begin{eqnarray}
 \rho_{\chi}(\tau)=\frac{1}{4\pi^2 a^4(\tau)}\int_0^{\infty}(|\bar{\chi}_k'|^2+\omega_k^2(\tau)|\bar{\chi}_k |^2)k^2dk,
 \end{eqnarray}
which is a divergent quantity. Thus, this quantity has to be renormalized, 
the most popular way to do it is to use the adiabatic regularization, which consists in subtracting the zero, second and fourth order adiabatic expressions of the energy density (see for instance \cite{Bunch}). In the appendix of \cite{ha} it has been shown that the leading term of the renormalized energy density of the produced particles
after the phase transition is given by 
\begin{eqnarray}
 \rho_{\chi}^{ren}(\tau)=\frac{1}{2\pi^2 a^4(\tau)}\int_0^{\infty} \omega_k(\tau)k^2|\beta_k|^2dk,
 \end{eqnarray}
and for our model, using cosmic time, we will have
\begin{eqnarray}
 \rho_{\chi}^{ren}(t)=\frac{m^6}{512\pi^2 m_{\chi}^2}\int_0^{\infty} \frac{x^2\sqrt{x^2\left(\frac{a_{kin}}{a(t)}  \right)^2+1}}{(x^2+1)^5}dx \left( \frac{a_{kin}}{a(t)} \right)^3.
 \end{eqnarray}

Note that the quantity  $\frac{a_{kin}}{a(t)}$ decreases very fast in the kination regime. Effectively, in this regime one has, 
$a(t)=a_{kin}\left(t/t_{kin}\right)^{1/3}$, 
with $t_{kin}=\frac{1}{3H_{kin}}$, then at time 
$\bar{t}=\frac{10^6}{3H_{kin}}\sim 10^{11} M_{pl}^{-1}$, one has $\frac{a_{kin}}{a(\bar{t})}\cong 10^{-2}$, which as we see is a very small time compared with the time the universe spends in  the kination regime. Then one can conclude that during  the kination phase
one has $\frac{a_{kin}}{a(t)}\ll 10^{-2}$. Hence, the renormalized energy density is approximately equal to
\begin{eqnarray}
 \rho_{\chi}^{ren}(t)=\frac{m^6}{512\pi^2 m_{\chi}^2}\int_0^{\infty} \frac{x^2}{(x^2+1)^5}dx \left( \frac{a_{kin}}{a(t)} \right)^3\cong
 10^{-5}\left(\frac{m}{m_{\chi}}\right)^2 m^4\left(\frac{a_{kin}}{a(t)} \right)^3.
 \end{eqnarray}

Note that this expression could be written as follows $\rho_{\chi}^{ren}(t)=m_{\chi}n_{\chi}(t)$, where
\begin{eqnarray}
n_{\chi}(t)=\frac{1}{2\pi^2 a^3(t)}\int_0^{\infty} k^2|\beta_k|^2dk,
\end{eqnarray}
has to be understood as the number density of produced particles at the phase transition.

We consider the decay  of the  $\chi$-field in fermions ($\chi\rightarrow \psi\bar\psi$) via a  Yukawa-type interaction  $h\psi\bar\psi\chi$, giving rise to the decay rate $\bar{\Gamma}=\frac{h^2 m_{\chi}}{8\pi}$ \cite{lindebook},
which will be finished  when $\bar{\Gamma}\sim H(t_{dec})\equiv H_{dec}$. First of all, we impose that the decay was before the end of kination, that means, before the equality between the energy density of the field and the one of the produced particles. Thus, for the universe staying in the kination regime we will have $H_{dec}=H_{kin}\left(\frac{a_{kin}}{a_{dec}} \right)^3\cong
  \sqrt{ \frac{1+\sqrt{3}}{6} }{m}\left(\frac{a_{kin}}{a_{dec}}\right)^3 $, and the corresponding energy densities will be
\begin{eqnarray}
\rho_{dec}\equiv \rho(t_{dec})=3\bar{\Gamma}^2M_{pl}^2, \quad \mbox{and} \quad  \rho_{\chi, dec}^{ren}\equiv \rho_{\chi}^{ren}(t_{dec})\sim 1.5\times 10^{-5}\left( \frac{m}{m_{\chi}} \right)^2 \frac{\bar{\Gamma}}{m}m^4.
\end{eqnarray}

On the other hand, from the condition  $\rho_{\chi, dec}^{ren} \leq \rho_{dec}$, one gets
\begin{eqnarray}\label{BOUND}
h^2\geq {4\pi}\times 10^{-5}\left( \frac{m}{m_{\chi}} \right)^3 \left(\frac{m}{M_{pl}}  \right)^2,
\end{eqnarray}
which for the value of the inflaton mass $m\cong 5\times 10^{-6}M_{pl}$, obtained from the theoretical and observational values of the power spectrum of scalar perturbations, and taking the bare mass of the quantum field  
as $m_{\chi}\cong 5\times 10^{-4}M_{pl}$, 
imposes a restriction on the coupling constant as $h\geq  6\times10^{-11}$. Moreover, one has to assume that the decay is after the beginning of the kination, which implies that $\bar{\Gamma}\leq H_{kin}$, obtaining $h^2\leq \frac{8\pi H_{kin}}{m_{\chi}}$, which for the values of $H_{kin}$ and $m_{\chi}$ gives another restriction as,
$ h\leq 4\times 10^{-1}$. Thus, we have obtained that the parameter $h$ is constrained as  $6\times 10^{-11}\leq h \leq 4\times 10^{-1}$.

Assuming instantaneous thermalization, 
the reheating temperature, i.e., the temperature of the universe when the relativistic plasma in thermal equilibrium starts to dominate, 
could be calculated as follows. The evolution of the energy density of the created particles and background respectively are,
\begin{eqnarray}
\rho_{\chi}^{ren}(t)=\rho_{\chi, dec}^{ren}\left( \frac{a_{dec}}{a(t)} \right)^4 ,\qquad 
\rho(t)=\rho_{dec}\left( \frac{a_{dec}}{a(t)} \right)^6, 
\end{eqnarray}
which tells us that at the time when  
the kination phase ends, namely $t_r$, i.e., when 
$\rho_{r}\equiv \rho(t_{r})=\rho_{\chi}^{ren}(t_r)\equiv \rho_{\chi, r}^{ren} $, 
the reheating temperature can be  calculated as follows:

Since at  $t_{r}$ we will have,
$ \left( \frac{a_{dec}}{a_{r}} \right)^2=\frac{\rho_{\chi, dec}^{ren}}{\rho_{dec}}$, thus, the reheating temperature takes the form 
\begin{eqnarray}\label{reheating1}
 T_R=  \left(\frac{30}{\pi^2 g_*} \right)^{1/4}  (\rho_{\chi, r}^{ren})^{1/4}
  =\left(\frac{30}{\pi^2 g_*} \right)^{1/4}(\rho_{\chi, dec}^{ren})^{1/4}\sqrt{\frac{\rho_{\chi, dec}^{ren}}{\rho_{dec}}} 
\nonumber \\ \cong 
2\times 10^{-4} g_*^{-1/4}\left(\frac{m}{m_{\chi}}  \right)^{3/2}\left(\frac{m}{\bar\Gamma}  \right)^{1/4}\left(\frac{m}{M_{pl}}  \right)^2 M_{pl}~.
\end{eqnarray}
This reheating temperture [i.e., eqn. (\ref{reheating1})], using the above values of the inflaton mass $m$ ($\cong 5\times 10^{-6}M_{pl}$) and $m_{\chi}$ ($\cong 5\times 10^{-4}M_{pl}$), can be approximated as 

\begin{eqnarray}\label{reheatingtemperature}
T_R\cong 3.5\times 10^{-18} h^{-1/2}g_*^{-1/4} M_{pl}\cong 8 h^{-1/2} g_*^{-1/4}\mbox{ GeV}.
\end{eqnarray}

\begin{remark}
The thermalization is nearly an instantaneous process. First of all we write the energy density of the decay products, which are very light particles, as
\begin{eqnarray*}
\rho_{\chi}^{ren}(t)=RH_{dec}^4\left(\frac{a_{dec}}{a(t)}\right)^4, 
\end{eqnarray*}
where we have introduced the notation $R=\frac{\rho_{\chi, dec}^{ren}}{H_{dec}^4}$. Now, following the reasoning of \cite{pv}, the decay products have a typical energy of the form  $\bar{\epsilon}\sim H_{end}\left(\frac{a_{dec}}{a(t)}\right)$, and its number density is  $n\sim R\bar{\epsilon}^3$.

Taking into account that,   if the particles interact
by the exchange of gauge bosons and establish thermal
equilibrium among the fermions and gauge bosons,
 the interaction  rate will be $n\sigma$, where the cross section is given by $\sigma\sim \frac{\alpha^2}{\bar{\epsilon}^2}$, with the coupling constant satisfying the inequality $10^{-2}\leq \alpha\leq 10^{-1}$. The thermal equilibrium will be accomplished when the interaction rate becomes comparable to the
Hubble parameter $H=H_{dec}\left(\frac{a_{dec}}{a(t)}\right)^3$, which happens when $\left(\frac{a_{dec}}{a_{th}}\right)^2=R\alpha^2$, where the subscript ``~${th}$'' attached to any quantity refers to its value at the time when the thermal equilibrium has been established.

In fact, one can  calculate the scale factor at $t = t_{rh}$ as follows:
\begin{eqnarray*}
 \left( \frac{a_{dec}}{a_{r}} \right)^2=\frac{\rho_{\chi, dec}^{ren}}{\rho_{dec}}=\frac{H_{dec}^2R}{3M_{pl}^2}=\frac{H_{dec}^2}{3M_{pl}^2\alpha^2}\left( \frac{a_{dec}}{a_{th}} \right)^2 
 \Longrightarrow a^2_{th}=\frac{H_{dec}^2}{3M_{pl}^2\alpha^2}a_r^2. \end{eqnarray*}

Now, since $H_{dec}\leq H_{kin}$ one has
\begin{eqnarray*}
a^2_{th}\leq \frac{H_{kin}^2}{3M_{pl}^2\alpha^2}a_r^2\cong \frac{3}{\alpha^2}\times 10^{-12}a_r^2\leq 3\times 10^{-10}a_r^2\Longrightarrow a_{th}\ll a_r,
\end{eqnarray*}
which means that the thermal equilibrium occurs well before the equality between the energy density of the scalar field and the one of the decay products, and thus,
one could safely assume an instantaneous thermalization.

\end{remark}

Now we assume that the decay of the $\chi$-field is after the end of kination. 
Since the decay is after  $t_r$, one has to impose $\bar{\Gamma}\leq H_{r}$. Taking this into account, one has 
\begin{eqnarray}\label{31}
H^2_{r}=\frac{2\rho_{r}}{3M_{pl}^2}\quad \mbox{and} \quad \rho_{r}=\rho_{kin}\left( \frac{a_{kin}}{a_{r}} \right)^6=3H^2_{kin}M_{pl}^2\Theta^2,
\end{eqnarray}
where we have introduced the so-called {\it heating efficiency} defined as $\Theta\equiv \frac{\rho^{ren}_{\chi, kin}}{\rho_{kin}}$. Consequenlty, from eqn. (\ref{31}), one can easily have $H_{r}=\sqrt{2}H_{kin}\Theta$, and thus, one obtains that the parameter $h$ has to be
very small satisfying $h\leq 6\times 10^{-11}$.  Assuming once again the instantaneous thermalization, the reheating temperature (i.e., the temperature of the universe when the thermalized plasma starts to dominate) will be
\begin{eqnarray}
T_R=\left( \frac{30}{\pi^2 g_*} \right)^{1/4}(\rho_{\chi, dec}^{ren})^{1/4}= \left( \frac{90}{\pi^2 g_*} \right)^{1/4}\sqrt{\bar{\Gamma}M_{pl}},
\end{eqnarray}
where we have used that after  $t_r$, the energy density of the produced particles dominates the energy density of the inflaton field. 
Then, we will have 

\begin{eqnarray}
T_R\cong 7\times 10^{-3} hg_*^{-1/4} M_{pl}\Longrightarrow  T_R\leq 4.2 g_*^{-1/4}10^{-13} M_{pl}\cong  10^{6} \mbox{GeV}.
\end{eqnarray}

Consequently,  assuming that the BBN epoch occurs at the $1$ MeV regime, this constrains the value of $h$ residing in the interval $10^{-19}\leq h\leq 6\times 10^{-11}$, and the reheating temperature, depending on $h$,  will be  in the TeV, GeV  or in the MeV regime.

At the end of this Section we need to show that the time $\bar{t}\sim 10^{11} M_{pl}^{-1}$ at which $\frac{a_{kin}}{a(t)}\cong 10^{-2}$ is very small compared with the time that kination lasted. To simplify, we assume that the decay is before  the end of kination although the reasoning is similar in the other situation. In that case we have
\begin{eqnarray}
\Theta=\left( \frac{a_{kin}}{a(t_{r})} \right)^3\sim \frac{1}{H_{kin} t_{r}}
\Longrightarrow t_{r}\sim \frac{1}{H_{kin}\Theta}\sim 10^{26} M_{pl}^{-1}.
\end{eqnarray}

\section{BBN constraints  coming from the production of  Gravitational Waves}
\label{sec-BBN}

This section is devoted to present the bounds on the proposed improved version of the  {\it quintessential inflationary} model using the Big Bang  Nucleosynthesis (BBN) where we explicitly use the BBN constraints from the logarithmic spectrum of GWs and consequently the BBN bounds from the overproduction of GWs.

\subsection{BBN constraints from the logarithmic spectrum of GWs}
\label{subsec-gw1}

It is well known that during inflation, the GWs  are produced (known as the primordial GWs, in short PGWs), and in the post-inflationary, i.e., during kination, the logarithmic spectrum of GWs, namely,
$\Omega_{GW}$ defined as $\Omega_{GW}\equiv \frac{1}{\rho_c}\frac{d\rho_{GW}(k)}{d\ln k }$ (where $\rho_{GW}(k)$ is the energy density spectrum of the produced GWs; $\rho_c=3H_0^2M_{pl}^2$, where $H_0$ is the present value of the Hubble parameter is the so-called {\it critical density}) scales as $k^2$ \cite{rubio}, producing a spike in the spectrum of GWs at high frequencies. Then in order that GWs do not destabilize the BBN, the following bound must be imposed  (see Section 7.1 of \cite{maggiore})
\begin{eqnarray}\label{integral}
I\equiv h_0^2\int_{k_{BBN}}^{k_{end}} \Omega_{GW}(k) d \ln k \leq 10^{-5},
\end{eqnarray}
where $h_0\cong 0.678$ parametrizes the experimental uncertainty to determine the current value of the Hubble constant and $k_{BBN}$, $k_{end}$ are respectively the momenta associated to the horizon scale at the BBN and at the end of inflation. As it has been  shown in \cite{Giovannini1} that the main contribution of this integral (\ref{integral}) comes from the modes that left the Hubble radius before the inflationary epoch and finally re-enters during the inflation, that means, for $k_{r}\leq k\leq k_{kin}$, where
$k_{r}=a_{r}H_{r}$ and $k_{kin}=a_{kin}H_{kin}$.  For these modes one can calculate the  logarithmic spectrum of GWs as in \cite{Giovannini} (see also \cite{rubio, Giovannini2, Giovannini3,Giovannini:2016vkr} where the  the graviton spectra
in quintessential models have been reassessed, in a model-independent way, using the numerical techniques)

\begin{eqnarray}\label{Omega}
\Omega_{GW}(k)=\tilde{\epsilon}\Omega_{\gamma}h^2_{GW} \left(\frac{k}{k_{r}}  \right)\ln^2\left(\frac{k}{k_{kin}}  \right),
\end{eqnarray}
where $h^2_{GW}=\frac{1}{8\pi}\left(\frac{H_{kin}}{M_{pl}}  \right)^2$
is the amplitude of the GWs; $\Omega_{\gamma}\cong 2.6\times 10^{-5} h_0^{-2}$, is the present density fraction of radiation; and the quantity $\tilde{\epsilon}$  which for the Standard Model of particle physics is approximately equal to $0.05$,  takes into account the variation of massless degrees of freedom between decoupling and thermalization (see \cite{rubio, Giovannini1} for more details). Now, plugging expression (\ref{Omega}) into (\ref{integral}), and disregarding the sub-leading logarithmic terms, one finds  
\begin{eqnarray}\label{constraintx}
 2\tilde{\epsilon}h_0^2\Omega_{\gamma}h^2_{GW}\left( \frac{k_{kin}}{k_{r}} \right)\leq 10^{-5}
 \Longrightarrow 10^{-2}\left(\frac{H_{kin}}{M_{pl}}  \right)^2 \left( \frac{k_{kin}}{k_{r}} \right)\leq 1.
  \end{eqnarray}

To calculate the ratio $k_{kin}/k_{r} $,  we  will have to study the following
 three different situations:
\begin{enumerate}
\item When the produced particles are very light and its energy density decays as $a^{-4}$.
In this case, as shown in  \cite{rubio},  one will have 
\begin{eqnarray}
\frac{k_{kin}}{k_{r}} =\frac{1}{\sqrt{2} \Theta},
\end{eqnarray}
where $\Theta$ is the {\it heating efficiency} introduced at the end of the previous Section. Thus,  the constraint (\ref{constraintx}) eventually directs 
\begin{eqnarray}
\Theta  \geq 7\times 10^{-3} \left(\frac{H_{kin}}{M_{pl}}  \right)^2.\end{eqnarray}

\item If the produced particles have heavy masses and their energy densities decay as $a^{-3}$, before their decays compared to the light particles. In this situation there are two sub-cases:
\begin{enumerate}
\item If the decay happens before the end of kination. For this sub-case, 
a simple calculation leads to 
\begin{eqnarray}
\frac{k_{kin}}{k_{r}} =\frac{1}{\sqrt{2} \Theta}\left( \frac{\bar{\Gamma}}{H_{kin}}  \right)^{1/3},
\end{eqnarray}
and consequently the constraint (\ref{constraintx}) becomes
\begin{eqnarray}
\Theta \left(\frac{H_{kin}}{\bar{\Gamma}}  \right)^{1/3}\geq 7\times 10^{-3} \left(\frac{H_{kin}}{M_{pl}}  \right)^2,
\end{eqnarray}
which applied to our model finally leads to
\begin{eqnarray}\label{2a}
h^2\leq 3\times 10^{-9} \left(\frac{m}{m_{\chi}} \right)^7\left(\frac{m}{H_{kin}} \right)^{11}.
\end{eqnarray}

\item When the decay happens after the end of kination.
In this case one has  
\begin{eqnarray}
\frac{k_{kin}}{k_{r}} =\frac{1}{\sqrt{2} \Theta^{2/3}},
\end{eqnarray}
and the constraint (\ref{constraintx}) leads to
\begin{eqnarray}\label{2b}
\Theta^{2/3}  \geq 7\times 10^{-3} \left(\frac{H_{kin}}{M_{pl}}  \right)^2.\end{eqnarray}

\end{enumerate}
\end{enumerate}

The first case, i.e., when the particles are very light, does not fit here, because we are dealing with heavy massive particles. In the case,  $2(a)$, i.e., when the decay is before the end of kination, the constraint (\ref{2a}) together with the the bound (\ref{BOUND}), coming from the imposition that the decay was before 
the end of kination, leads to the condition 
\begin{eqnarray}
\frac{16\pi}{3}\times 10^{-5}\left( \frac{m}{m_{\chi}} \right)^3 \left(\frac{m}{M_{pl}}  \right)^2
\leq 3\times 10^{-9} \left(\frac{m}{m_{\chi}} \right)^7\left(\frac{m}{H_{kin}} \right)^{11},
\end{eqnarray}
with for the values of $m$ and $H_{kin}$ 
 bounds the value of the mass 
of the quantum field to satisfy 
$m_{\chi}\leq 6\times 10^{-4} M_{pl}$, which is compatible with our choice $m_{\chi}= 5\times 10^{-4} M_{pl}$ to ensure that one can apply the WKB approximation. So,  in this case, the gravitational waves do not  affect the  BBN success. Moreover, for the value $m_{\chi}= 5\times 10^{-4} M_{pl}$  one obtains the bound
$ 6\times 10^{-11}\leq h\leq 10^{-10}$, and for these values, taking $g_*=107$,  the reheating temperature (\ref{reheatingtemperature}) is around $240$ TeV.

Finally, in the situation $2(b)$, i.e., when the decay occurs after the end of kination, the condition (\ref{2b}) bounds the value of the mass of the quantum field in order to satisfy  $m_{\chi}\leq  10^{-3} M_{pl}$, which is fulfilled for our choice, that means, the BBN constraint (\ref{integral}) is always overpassed when the decay occurs  after
the end of kination.

\subsection{BBN bounds from the overproduction of GWs}
\label{subsec-gw2}

The success of the BBN demands that \cite{dimopoulos}
\begin{eqnarray}\label{bbnconstraint}
\frac{\rho_{GW}(t_{reh})}{\rho_{\chi}^{ren}(t_{reh})}\leq 10^{-2},
\end{eqnarray}
where $t_{reh}$ is the reheating time and $\rho_{GW}(t)$ is the energy density of the GW produced  at the phase transition.  The value of the energy density of the GWs is 
usually taken to be, $\rho_{GW}(t)\cong 10^{-2} H^4_{kin} \left( \frac{a_{kin}}{a(t)} \right)^4$ (see for example \cite{pv, Damour}), although we have discussed this point later in {\it Appendix A}. 

Firstly, we assume that the decay occurs after  the end of kination,  and we calculate $\frac{\rho_{GW}(t_{r})}{\rho_{\chi, r}^{ren}}$.
Using equation (\ref{31}) and the fact that $\Theta=\left(a_{kin}/a_{r}  \right)^3$, one finds 
\begin{eqnarray}
\frac{\rho_{GW}(t_{r})}{\rho_{\chi, r}^{ren}}= \frac{1}{3}10^{-2} \left(\frac{H_{kin}}{M_{pl}}\right)^2\Theta^{-2/3}\cong 3\times 10^{-1},
\end{eqnarray}
which means that if the decay occurs before the end of kination, the constraint (\ref{bbnconstraint}) is never achieved, because, after the decay, the energy densities of the produced particles decrease as the one of the GWs.  Hence, the decay must occur after  $t_r$ and assuming once again the instantaneous thermalization,  the reheating time will coincide with the decay one. Then, we will have 
$\rho_{\chi, dec}^{ren}=3\bar{\Gamma}^2M_{pl}^2$, and since 
\begin{eqnarray}
H_{dec}=H_{r}\left( \frac{a_{r}}{a_{dec}} \right)^{3/2}\Longrightarrow \left( \frac{a_{r}}{a_{dec}} \right)^{3/2}=\frac{\bar\Gamma}{\sqrt{2}H_{kin}\Theta},
\end{eqnarray}
we have 
\begin{align}
\rho_{GW}(t_{dec})=\rho_{GW}(t_{r})\left( \frac{a_{r}}{a_{dec}} \right)^4= \rho_{GW}(t_{r})\left( \frac{\bar\Gamma}{\sqrt{2}H_{kin}\Theta}    \right)^{8/3}
=10^{-2} H^4_{kin}\Theta^{-4/3}\left( \frac{\bar\Gamma}{\sqrt{2}H_{kin} }  \right)^{8/3},
\end{align}
and thus,
\begin{eqnarray}
\frac{\rho_{GW}(t_{reh})}{\rho_{\chi}^{ren}(t_{reh})}\cong 10^{-4} \left(\frac{h}{\Theta}  \right)^{4/3}\frac{m_{\chi}^{2/3}H^{4/3}_{kin}}{M_{pl}^2}\cong 2\times 10^{13} h^{4/3},
\end{eqnarray}
meaning that the constraint (\ref{bbnconstraint}) is satisfied for $h\leq 3\times 10^{-12}$. Therefore, since in the previous Section \ref{sec-III}, we have showed that a  successful reheating where the decay occurring after the equality between the energy density of the scalar field and the one of the produced particles demands that $10^{-19}\leq h\leq 6\times 10^{-11}$, thus, one can conclude that in order to avoid problems in the BBN due to the overproduction of GW one has to choose $10^{-19}\leq h\leq 3\times 10^{-12}$. Moreover, this enables one to obtain a reheating temperature lower than $15$ TeV.

\section{Observational data and the results}
\label{sec-data+results}

In this section we present the observational constraints on the quintessence piece of the present model (\ref{PV}) using the latest astronomical datasets, namely, the cosmic microwave background radiation, baryon acoustic oscillations distance measurements, Pantheon sample from the Supernovae Type Ia, and the Hubble parameter measurements from the cosmic chronometers. In what follows we describe the observational datasets and the results.

\begin{itemize}

\item CMB: We use the cosmic microwave background (CMB) radiation from the Planck 2015 measurements. The CMB temperature and polarization anisotropies along with their cross-correlations from the Planck 2015 \cite{Adam:2015rua} have been employed in the analysis. Specifically, the combinations of high- and low-$\ell$ 
TT likelihoods in the multiple range $2\leq \ell \leq 2508$ as well as  the combinations  of the high- and low-$\ell$ polarization likelihoods \cite{Aghanim:2015xee} have been considered. 

\item BAO: Along with the CMB measurements, we consider the 
Baryon acoustic oscillation (BAO) distance  measurements from diverse 
astronomical missions, namely, 6dF
Galaxy Survey (6dFGS) \cite{Beutler:2011hx}; Main Galaxy Sample of Data Release 7 of Sloan Digital Sky Survey (SDSS-MGS) \cite{Ross:2014qpa}; CMASS and LOWZ samples from the latest Data Release 12
(DR12) of the Baryon Oscillation Spectroscopic Survey (BOSS) \cite{Gil-Marin:2015nqa}. The measurements by the above astronomical missions are as follows.  
From the 6dF Galaxy Survey (6dFGS), the redshift measurement is at $z_{\emph{\emph{eff}}}=0.106$
) \cite{Beutler:2011hx}; from the Main Galaxy Sample of Data Release 7 of SDSS-MGS is at  $z_{\emph{\emph{eff}}}=0.15$) \cite{Ross:2014qpa}; from the CMASS and LOWZ samples of the latest Data Release 12
(DR12) of BOSS are respectively at $z_{\mathrm{eff%
}}=0.57$ \cite{Gil-Marin:2015nqa} and at $z_{\mathrm{eff}}=0.32$) 
\cite{Gil-Marin:2015nqa}.

\item SNIa: In this work we take into account the latest compilation of the Supernovae Type Ia (SNIa) data known as {\it Pantheon sample} \cite{Scolnic:2017caz} consisting of 1048 data points in the redshift range $z \in [0.01, 2.3]$.

\item CC: We include the Hubble 
parameter measurements from the cosmic chronometers (CC) \cite{Moresco:2016mzx} comprising $30$ measurements in the redshift range $0< z < 2$ (see Table 4 of \cite{Moresco:2016mzx}). 
\end{itemize}
The observational constraints of the quintessence model given in eqn. (\ref{PV}) have been extracted using the  markov chain monte carlo package {\it cosmomc} \cite{Lewis:2002ah}  which  is equipped with an efficient convergence diagnostic by Gelman and Rubin \cite{Gelman-Rubin}. The code implements an efficient sampling to calculate the posterior distribution for each free parameter with the use of fast/slow parameter de-correlations \cite{Lewis:2013hha} (the code is publicly available at \url{http://cosmologist.info/cosmomc/}). Let us describe the observational constraints that we have from the potential (\ref{PV}). Let us  note that we have performed the analyses for three different cases of the parameter $M$, namely, $M = 1 $ GeV, $M = 20$ GeV and finally we allow $M$ to be a free parameter varying in the region $[-50, 50]$ aiming to see how the parameter $M$ is constrained for the present model.

\begin{figure}
\includegraphics[width=0.65\textwidth]{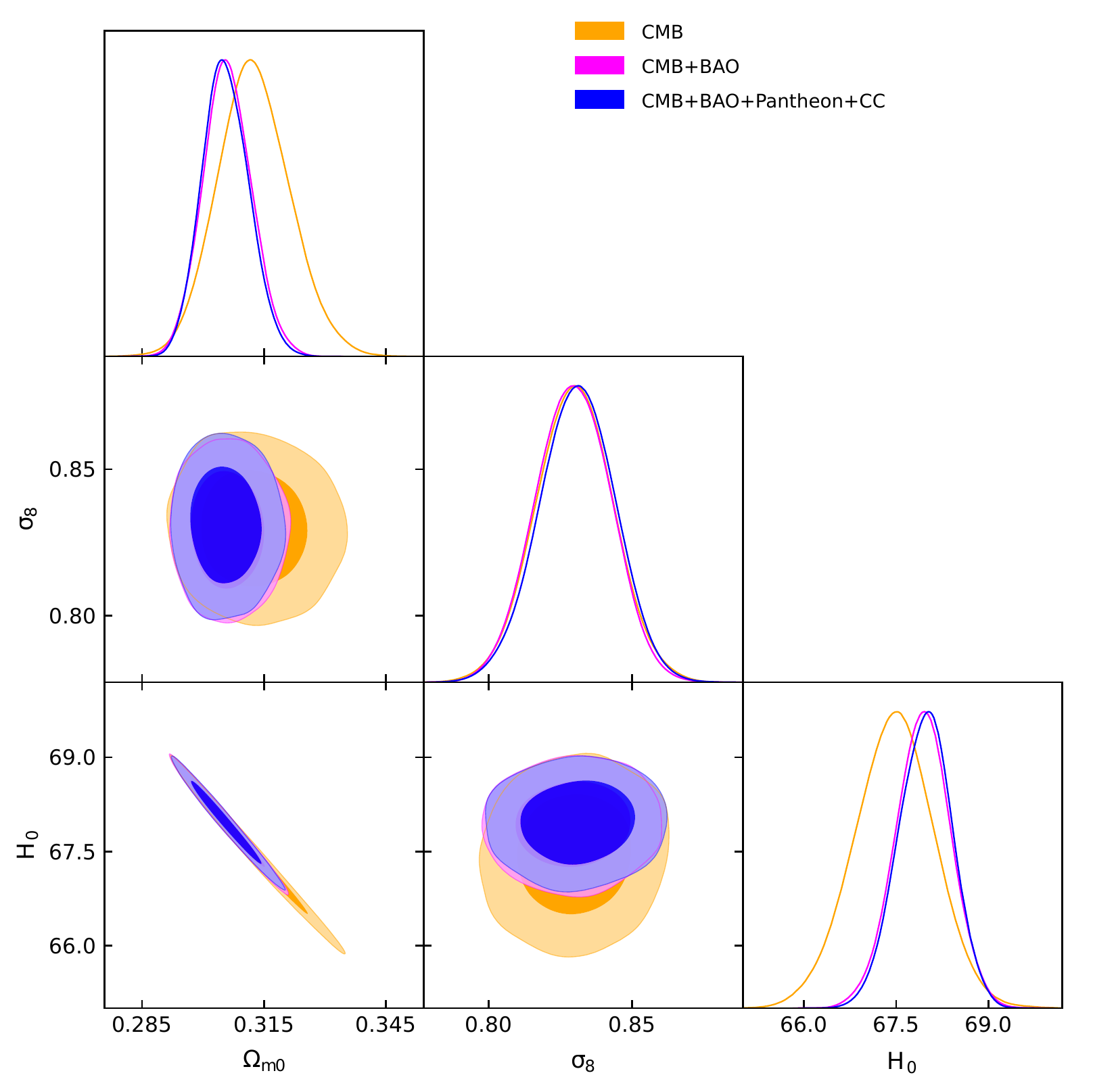}
\caption{68\% and 95\% confidence-level contour plots for several combinations of the model parameters for the quintessence potential (\ref{PV}) with fixed $M = 1$ GeV, using various combinations of the astronomical datasets. The plots also show the 1 dimensional marginalized posterior distributions for some model parameters as well.  }
\label{fig-model1}
\end{figure}

In Table \ref{tab:resultsI} we summarize the observational constraints on the quintessence  model for $M =  1$ GeV using various datasets, namely, CMB, CMB+BAO and CMB+BAO+Pantheon+CC. And in Fig. \ref{fig-model1} we show the one-dimensional marginalized posterior distributions for some derived parameters of the model as well as the two-dimensional contour plots between several combinations of the derived parameters at 68\% and 95\% confidence-level (CL).  We find that all three datasets return exactly similar constraints on the derived parameters. Additionally, one can notice that 
the Hubble constant at present, i.e., $H_0$, assumes similar values to Planck 2015 \cite{Ade:2015xua} and Planck 2018 \cite{Aghanim:2018eyx}, and thus, looking at the estimation of $H_0$ by Riess et al. 2016 \cite{Riess:2016jrr} reporting $H_0 = 73.24 \pm 1.74$ km s$^{-1}$ Mpc$^{-1}$ and Riess et al. 2018 reporting $H_0 =  73.48 \pm 1.66$ km/s/Mpc \cite{R18} (also see Ref. \cite{Birrer:2018vtm} for similar constraint on $H_0$ recently reported), one can see that the tension on $H_0$ is still existing in the improved version of the Peebles-Vilenkin potential. However, we find some interesting properties that are believed to represent the generic nature of the scalar field potentials. Looking at Fig. \ref{fig-model1}, we see that a strong negative correlation is present between the parameters $H_0$ and $\Omega_{m0}$. However, concerning the $H_0, \sigma_8$ parameters, we find that the contour of these two parameters is almost horizontal (see Fig. \ref{fig-model1}), hence, showing no correlations between them. Similarly, the contour of the parameters $(\Omega_{m0}, \sigma_8)$ is exactly vertical as seen from Fig. \ref{fig-model1} and thus, we do not find any kind of correlations between these parameters also. Exactly similar conclusions have been noticed in a class of quintessence models \cite{Yang:2018xah}.

\begin{figure}
\includegraphics[width=0.6\textwidth]{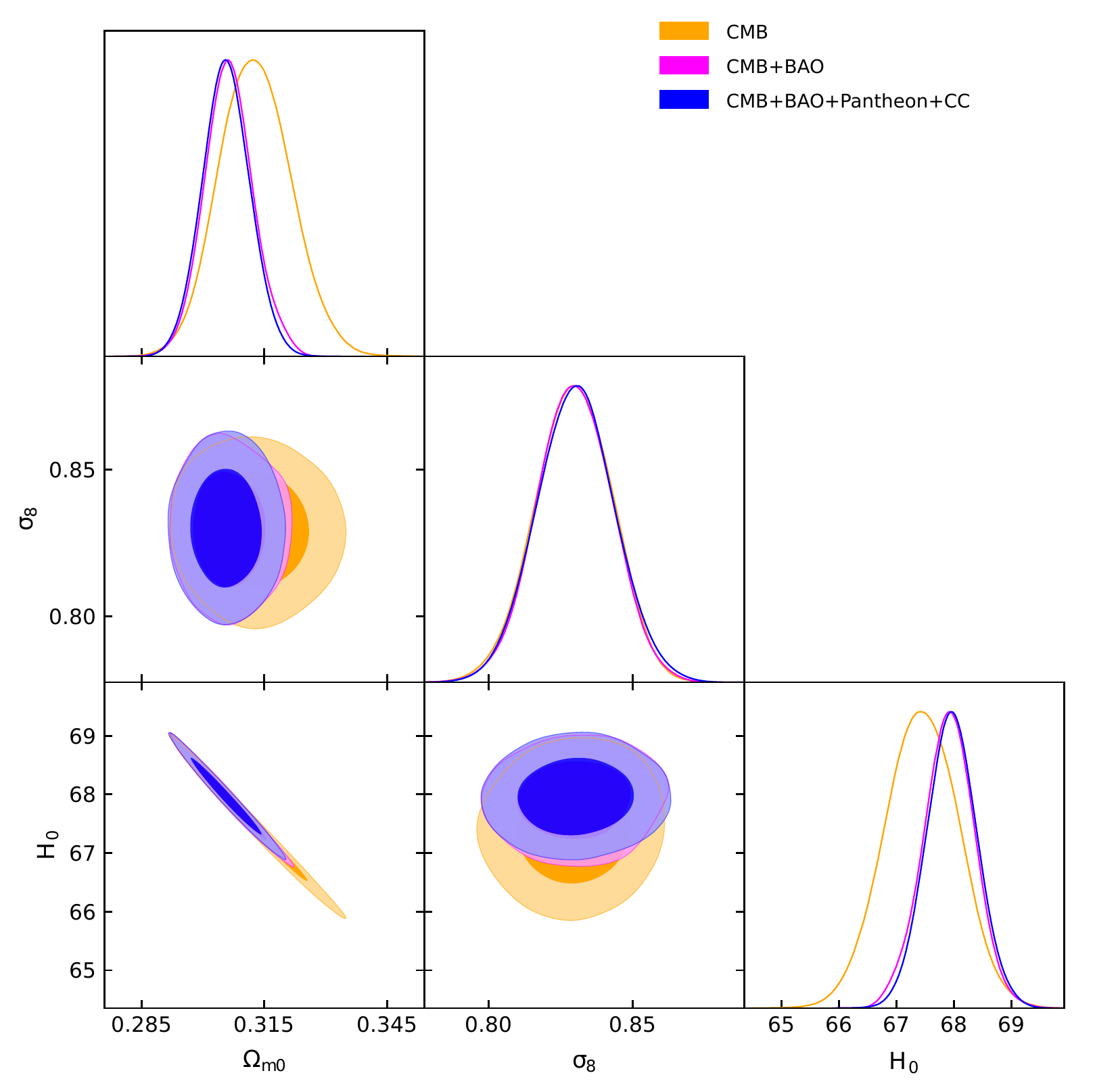}
\caption{68\% and 95\% confidence-level contour plots for several combinations of the model parameters for the potential (\ref{PV}) with fixed $M = 20$ GeV, using various combinations of the astronomical datasets. The plots also show the 1 dimensional marginalized posterior distributions for some model parameters as well. }
\label{fig-model2}
\end{figure}

We then fix the value of $M$ to $20$ GeV and constrain the quintessence piece of the improved quintessential inflationary model (\ref{PV}) using the same datasets employed in the previous case, that means, CMB, CMB+BAO and CMB+BAO+Pantheon+CC. The results are summarized in Table \ref{tab:resultsII} and the corresponding graphical distributions including the one-dimensional posterior distributions and the two-dimensional contour plots at 68\% and 95\% CL, are displayed in Fig. \ref{fig-model2}.  Our analyses report that similar to the previous case with $M =1$ Gev, here the constraints for all the combinations are almost similar, and moreover, 
although we expected some differences between the cosmological constraints for different values of the $M$ parameter, but that does not happen actually. As usual we find that the correlations between the parameters observed for the case with $M =1 $ GeV exist for this model too, and in addition to that, the tension on $H_0$ is still persisting. 
\begingroup                                                                                                                     
\squeezetable                                                                                                                   
\begin{center}                                                                                                                  
\begin{table}                                                                                                                   
\begin{tabular}{ccccccccccccccccc}
\hline \hline                                                                                                                                                                                                                                 
Parameters & CMB & CMB+BAO & CMB+BAO+Pantheon+CC\\ \hline
\hline
$\Omega_c h^2$ & $    0.1193_{-    0.0014-    0.0028}^{+    0.0015+    0.0029}$ & $    0.1182_{-    0.0010-    0.0019}^{+    0.0010+    0.0020}$ & $    0.1182_{-    0.0009-    0.0018}^{+    0.0010+    0.0019}$ \\

$\Omega_b h^2$ & $    0.02226_{-    0.00015-    0.00032}^{+    0.00015+    0.00031}$ & $    0.02232_{-    0.00014-    0.00027}^{+    0.00014+    0.00027}$ & $    0.02233_{-    0.00014-    0.00026}^{+    0.00014+    0.00027}$ \\

$100\theta_{MC}$ & $    1.04077_{-    0.00032-    0.00064}^{+    0.00032+    0.00065}$ & $    1.04090_{-    0.00030-    0.00059}^{+    0.00030+    0.00058}$ &  $    1.04091_{-    0.00029-    0.00061}^{+    0.00029+    0.00058}$ \\

$\tau$ & $    0.081_{-    0.017-    0.034}^{+    0.017+    0.034}$ & $    0.085_{-    0.016-    0.031}^{+    0.016+    0.031}$ & $    0.087_{-    0.016-    0.032}^{+    0.016+    0.031}$ \\

$n_s$ & $    0.9661_{-    0.0046-    0.0092}^{+    0.0050+    0.0091}$ & $    0.9688_{-    0.0038-    0.0074}^{+    0.0038+    0.0074}$ & $    0.9691_{-    0.0036-    0.0071}^{+    0.0036+    0.0070}$ \\

${\rm{ln}}(10^{10} A_s)$ & $    3.094_{-    0.033-    0.068}^{+    0.034+    0.066}$ & $    3.100_{-    0.032-    0.062}^{+    0.032+    0.061}$ & $    3.104_{-    0.032-    0.063}^{+    0.032+    0.060}$ \\

$\Omega_{m0}$ & $    0.313_{-    0.009-    0.017}^{+    0.009+    0.018}$ &  $    0.306_{-    0.006-    0.012}^{+    0.006+    0.012}$ & $    0.306_{-    0.006-    0.011}^{+    0.006+    0.012}$ \\

$\sigma_8$ & $    0.830_{-    0.013-    0.026}^{+    0.013+    0.026}$ &  $    0.829_{-    0.013-    0.026}^{+    0.013+    0.025}$ & $    0.831_{-    0.013-    0.026}^{+    0.013+    0.025}$  \\

$H_0$ & $   67.47_{-    0.65-    1.31}^{+    0.65+    1.27}$ & $   67.93_{-    0.45-    0.92}^{+    0.44+    0.89}$ & $   67.97_{-    0.44-    0.88}^{+    0.43+    0.84}$  
\\
\hline\hline                                                                                                                    
\end{tabular}                                                                                                                                                                                                                                      
\caption{68\% and 95\% confidence-level constraints on the quintessence potential (\ref{PV}) that means, the model (\ref{PV}) for $\varphi\geq -M_{pl}$, using various combinations of the astronomical datasets have been presented for fixed $M = 1$ GeV. Let us note that here $\Omega_{m0}$ is the present value of the total matter density $\Omega_m = \Omega_b +\Omega_c$; $H_0$, the present value of the Hubble constant, is in the units of km/Mpc/sec, and $\sigma_8$ is the amplitude of the matter power spectrum.  }\label{tab:resultsI}                                                                                                   
\end{table}                                                                                                                     
\end{center}                                                                                                                    
\endgroup
\begin{figure}
\includegraphics[width=0.6\textwidth]{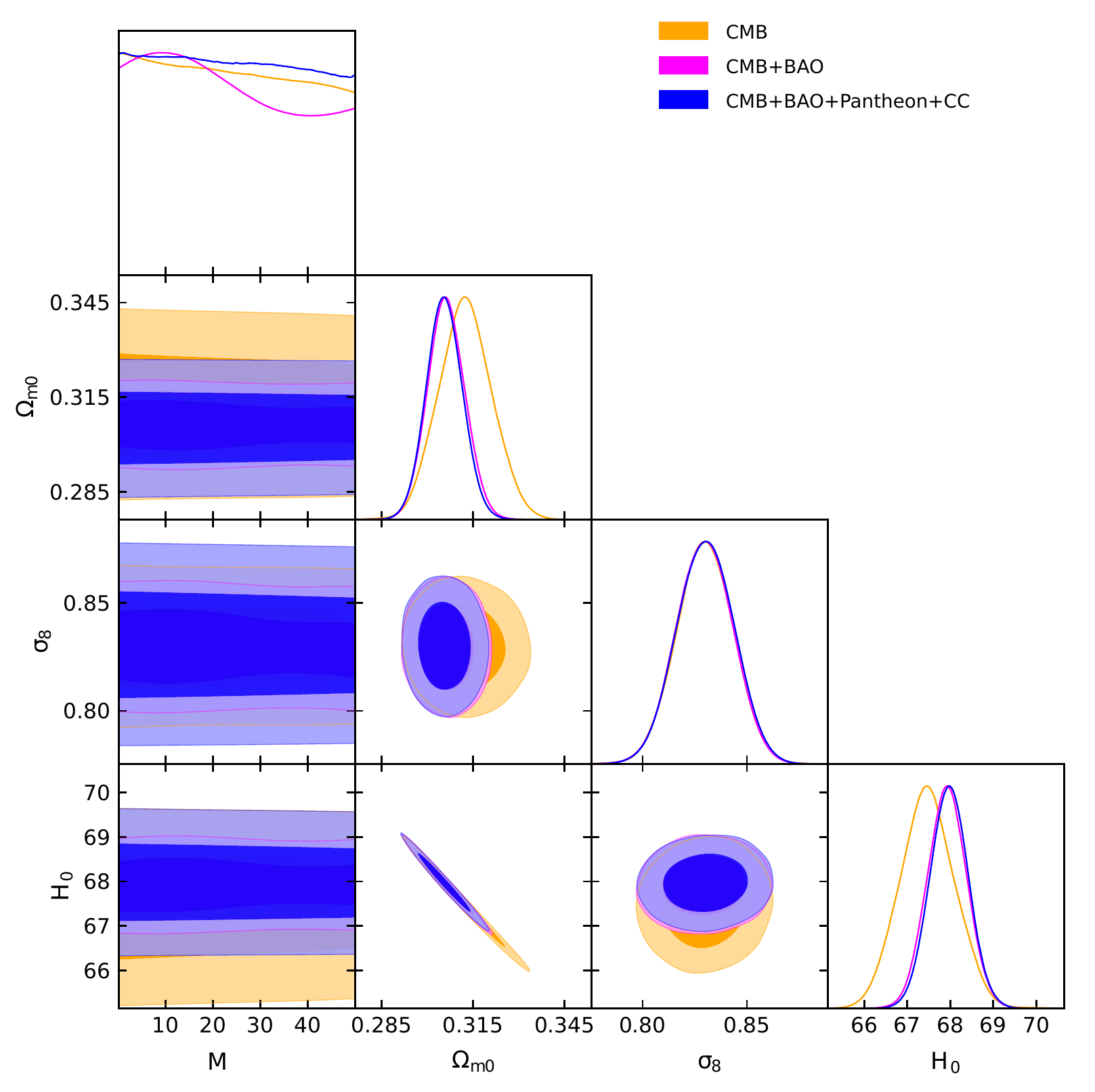}
\caption{68\% and 95\% confidence-level contour plots for several combinations of the model parameters for the potential (\ref{PV}) for varying $M$, using various combinations of the astronomical datasets. The plots also show the 1 dimensional marginalized posterior distributions for some model parameters as well. }
\label{fig-model3}
\end{figure} 
                                                                                                                       
Finally, we allow $M$ to be a free parameter for the quintessence piece of the model (\ref{PV}) and constrain this scenario with the same datasets used for the previous two models. The reuslts are summarized in Table \ref{tab:resultsIII} and the corresponding graphical distributions are shown in Fig. \ref{fig-model3}. We report that the current data cannot constrain $M$. It might be perhaps interesting to note that, although $M$ is kept free in this analysis, but, the presence/absece of correlations between the parameters that were observed for fixed values of $M$, do not change for free $M$ too.       
\begingroup                                                                                                                     
\squeezetable                                                                                                                   
\begin{center}                                                                                                                  
\begin{table}                                                                                                                   
\begin{tabular}{ccccccccccccccccc}
\hline \hline                                                                                                                                                                                                                                 
Parameters & CMB & CMB+BAO & CMB+BAO+Pantheon+CC\\ \hline

$\Omega_c h^2$ & $    0.1193_{-    0.0014-    0.0027}^{+    0.0014+    0.0029}$ & $    0.1183_{-    0.0010-    0.0020}^{+    0.0010+    0.0021}$ & $    0.1182_{-    0.0010-    0.0019}^{+    0.0010+    0.0019}$ \\

$\Omega_b h^2$ & $    0.02225_{-    0.00015-    0.00031}^{+    0.00015+    0.00030}$ & $    0.02232_{-    0.00014-    0.00028}^{+    0.00014+    0.00028}$ & $    0.02233_{-    0.00014-    0.00027}^{+    0.00014+    0.00027}$ \\

$100\theta_{MC}$ & $    1.04075_{-    0.00033-    0.00066}^{+    0.00033+    0.00063}$ & $    1.04090_{-    0.00031-    0.00060}^{+    0.00030+    0.00062}$ & $    1.04091_{-    0.00030-    0.00058}^{+    0.00030+    0.00059}$ \\

$\tau$ & $    0.080_{-    0.017-    0.035}^{+    0.018+    0.033}$ & $    0.085_{-    0.016-    0.032}^{+    0.016+    0.031}$ & $    0.086_{-    0.016-    0.032}^{+    0.016+    0.032}$  \\

$n_s$ & $    0.9658_{-    0.0045-    0.0088}^{+    0.0045+    0.0088}$ & $    0.9686_{-    0.0037-    0.0078}^{+    0.0038+    0.0073}$ & $    0.9689_{-    0.0037-    0.0072}^{+    0.0037+    0.0075}$ \\

${\rm{ln}}(10^{10} A_s)$ & $    3.093_{-    0.033-    0.067}^{+    0.034+    0.063}$ & $    3.100_{-    0.031-    0.064}^{+    0.032+    0.061}$ & $    3.102_{-    0.033-    0.065}^{+    0.033+    0.063}$ \\

$\Omega_{m0}$ & $    0.313_{-    0.009-    0.016}^{+    0.008+    0.018}$ & $    0.306_{-    0.006-    0.012}^{+    0.006+    0.013}$ & $    0.306_{- 0.006-    0.011}^{+    0.006+    0.012}$  \\

$\sigma_8$ & $    0.830_{-    0.013-    0.026}^{+    0.013+    0.025}$ & $    0.830_{-    0.013-    0.026}^{+    0.013+    0.025}$ &  $    0.830_{- 0.013-    0.026}^{+    0.013+    0.026}$ \\

$H_0$ & $   67.44_{-    0.63-    1.25}^{+    0.63+    1.24}$ & $   67.91_{-    0.44-    0.93}^{+    0.44+    0.88}$ & $   67.96_{-    0.44-    0.86}^{+    0.44+    0.87}$  \\
\hline \hline
\end{tabular}                                                                                                                   
\caption{68\% and 95\% confidence-level constraints on the quintessence potential (\ref{PV}) that means, the model (\ref{PV}) for $\varphi\geq -M_{pl}$, using various combinations of the astronomical datasets have been presented for fixed $M = 20$ GeV. Let us note that here $\Omega_{m0}$ is the present value of the total matter density $\Omega_m = \Omega_b +\Omega_c$, $H_0$, the present value of the Hubble constant, is in the units of km/Mpc/sec, and $\sigma_8$ is the amplitude of the matter power spectrum.   }\label{tab:resultsII}                                                                                                   
\end{table}                                                                                                                     
\end{center}                                                                                                                    
\endgroup   
\begingroup                                                                                                                     
\squeezetable                                                                                                                   
\begin{center}                                                                                                                  
\begin{table}                                                                                                                   
\begin{tabular}{cccccccccccccc}
\hline\hline                                                                                                                                                                                                                                
Parameters & CMB & CMB+BAO & CMB+BAO+Pantheon+CC\\ \hline

$\Omega_c h^2$ & $    0.1193_{-    0.0014-    0.0027}^{+    0.0014+    0.0027}$ & $    0.1183_{-    0.0010-    0.0020}^{+    0.0010+    0.0020}$  & $    0.1182_{-    0.0010-    0.0019}^{+    0.0010+    0.0019}$  \\

$\Omega_b h^2$ & $    0.02226_{-    0.00015-    0.00030}^{+    0.00015+    0.00030}$ & $    0.02232_{-    0.00014-    0.00026}^{+    0.00014+    0.00027}$ & $    0.02233_{-    0.00014-    0.00027}^{+    0.00014+    0.00027}$ \\

$100\theta_{MC}$ & $    1.04077_{-    0.00032-    0.00064}^{+    0.00032+    0.00062}$ & $    1.04089_{-    0.00033-    0.00059}^{+    0.00030+    0.00061}$ & $    1.04091_{-    0.00030-    0.00060}^{+    0.00030+    0.00058}$ \\

$\tau$ & $    0.081_{-    0.017-    0.033}^{+    0.017+    0.033}$ & $    0.085_{-    0.017-    0.033}^{+    0.017+    0.032}$ & $    0.086_{- 0.017-    0.033}^{+    0.017+    0.032}$ \\

$n_s$ & $    0.9661_{-    0.0045-    0.0087}^{+    0.0045+    0.0090}$ & $    0.9687_{-    0.0038-    0.0075}^{+    0.0038+    0.0075}$ & $    0.9690_{- 0.0038-    0.0076}^{+    0.0037+    0.0074}$ \\

${\rm{ln}}(10^{10} A_s)$ & $    3.094_{-    0.034-    0.066}^{+    0.034+    0.065}$ & $    3.100_{-    0.033-    0.065}^{+    0.033+    0.062}$ & $    3.102_{-    0.033-    0.064}^{+    0.033+    0.063}$ \\

$M$ & ${\rm unconstrained}$ & ${\rm unconstrained}$ & ${\rm unconstrained}$  \\

$\Omega_{m0}$ & $    0.312_{-    0.009-    0.017}^{+    0.009+    0.017}$ & $    0.306_{-    0.006-    0.012}^{+    0.006+    0.012}$  & $    0.306_{-    0.006-    0.011}^{+    0.006+    0.012}$ \\

$\sigma_8$ & $    0.830_{-    0.013-    0.026}^{+    0.013+    0.026}$ & $    0.830_{-    0.013-    0.026}^{+    0.013+    0.026}$ & $    0.830_{-    0.014-    0.026}^{+    0.014+    0.026}$ \\

$H_0$ & $   67.47_{-    0.63-    1.21}^{+    0.64+    1.23}$ & $   67.92_{-  0.46-    0.88}^{+    0.45+    0.90}$ & $   67.97_{-    0.44-    0.86}^{+    0.43+    0.88}$ \\
\hline\hline      
\end{tabular}                                                                                                                   
\caption{68\% and 95\% confidence-level constraints on the quintessence potential (\ref{PV}) that means, the model (\ref{PV}) for $\varphi\geq -M_{pl}$, using various combinations of the astronomical datasets have been presented for varying $M$. Let us note that here $\Omega_{m0}$ is the present value of the total matter density $\Omega_m = \Omega_b +\Omega_c$, $H_0$, the present value of the Hubble constant, is in the units of km/Mpc/sec, and $\sigma_8$ is the amplitude of the matter power spectrum.  }\label{tab:resultsIII}                                                                                                   
\end{table}                                                                                                                     
\end{center}                                                                                                                    
\endgroup
\begin{figure}
\includegraphics[width=0.3\textwidth]{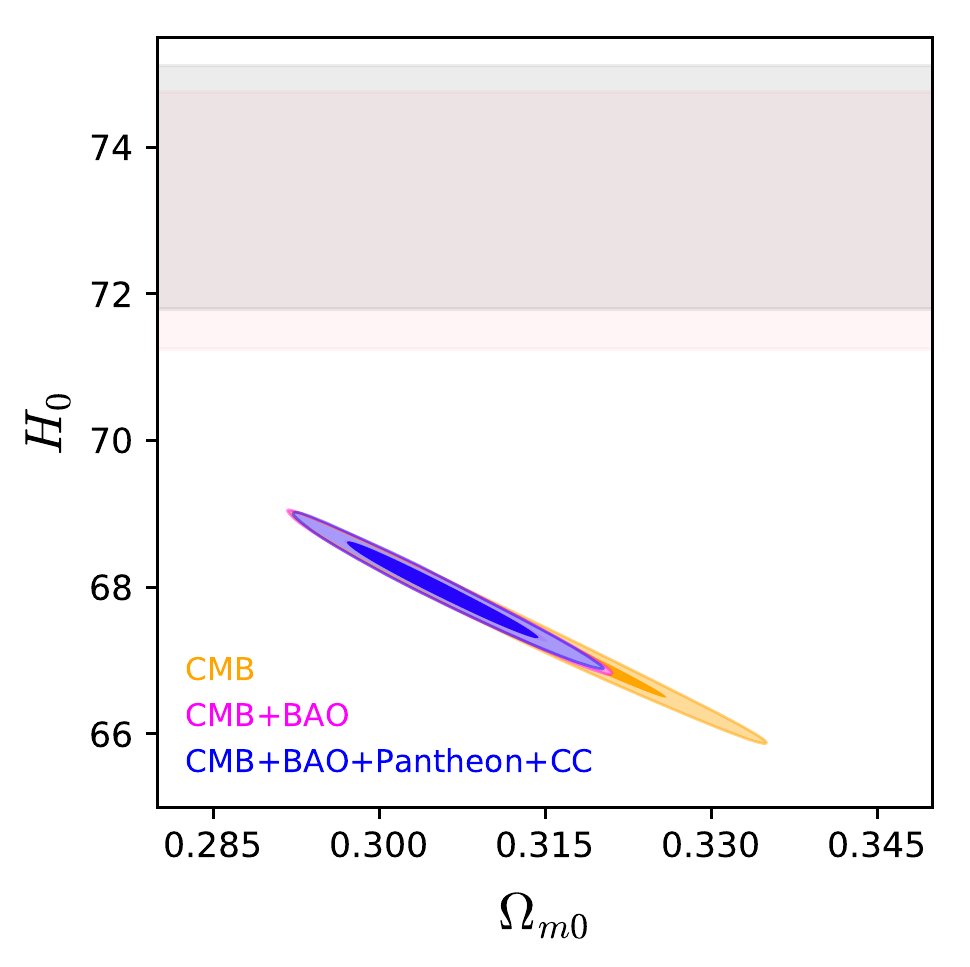}
\includegraphics[width=0.3\textwidth]{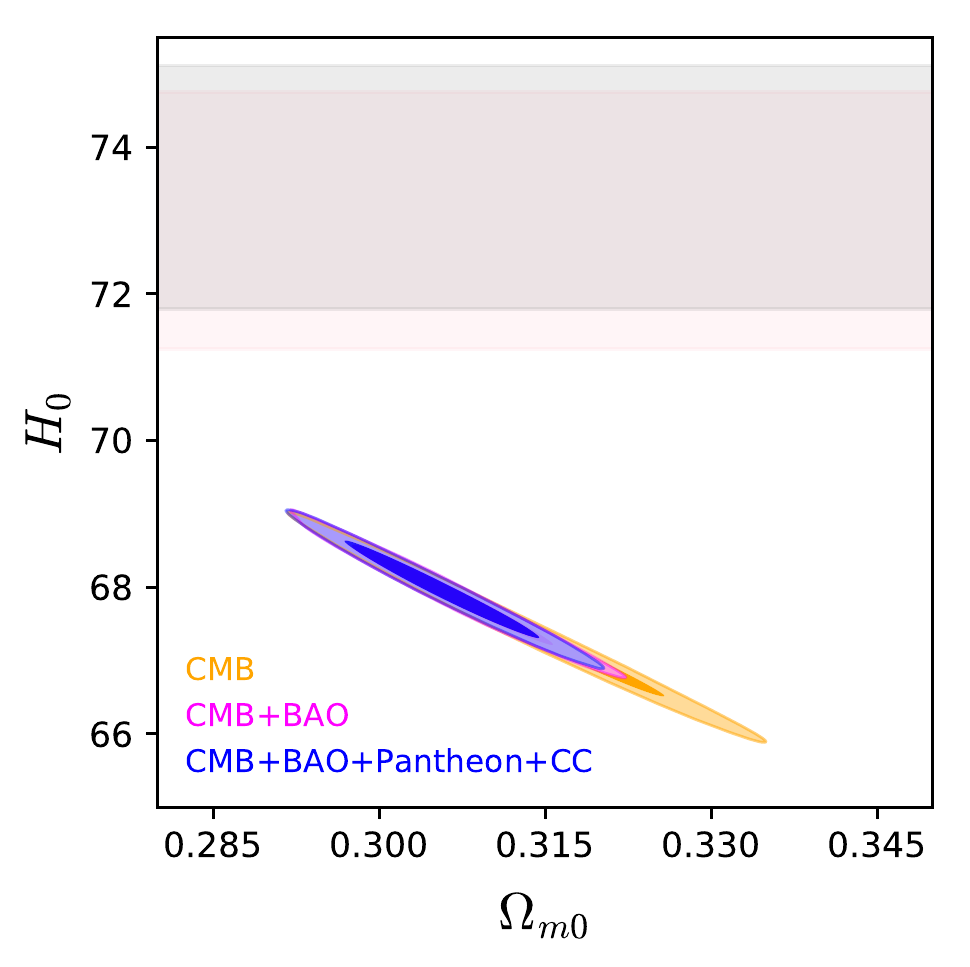}
\includegraphics[width=0.3\textwidth]{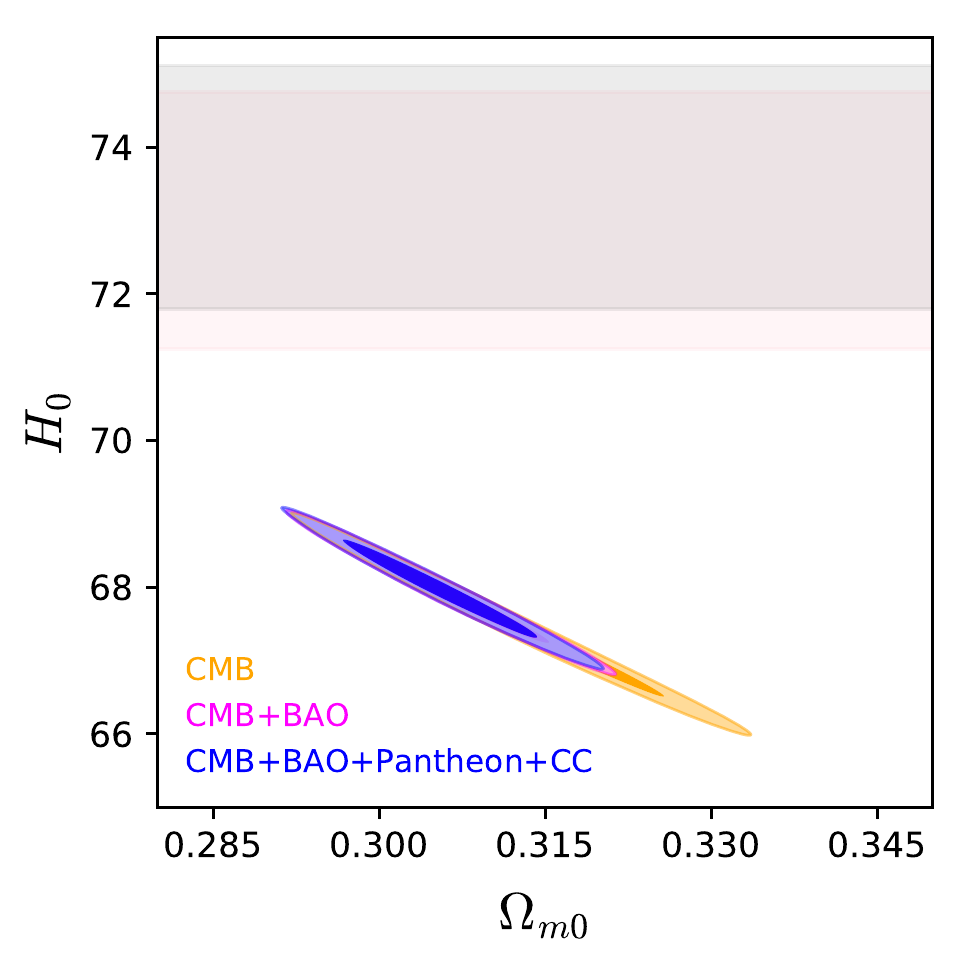}
\caption{Figures showing the tension on $H_0$ for the quintessence piece of the model (\ref{PV}) compared to the estimations by Riess et al 2016 \cite{Riess:2016jrr} (pink shaded region) and Riess et al 2018 \cite{R18} (grey shaded region) for $M=1 $ GeV (left panel), $M = 20$ GeV (middle panel) and $M$ treated as a free parameter (right panel) for all the observational datasets. }
\label{fig-R16-R18}
\end{figure}

\section{Concluding remarks}
\label{sec-conclu}

The quest of a unified theory connecting both early- and late- accelerated expansions is one of the biggest challenges for modern cosmology but is always enthralling. The theory of {\it quintessential inflation} is an attempt of such a unified theory that has been found quite impressive according to the investigations performed in the last couple of years. The well known model in this category is the Peebles-Vilenkin potential \cite{pv} in which the inflationary piece of the model is described by an quartic potential; however, for this model, the theoretical values of  scalar spectral index ($n_s$) and the ratio of tensor to scalar perturbations ($r$) do not enter into the marginalized  joint confidence contour in the plane  $(n_s,r)$ at $2\sigma$ CL, without the presence of running \cite{hap}. Since for this quintessential inflationary model, the number of $e$-folds is greater compared to what we find for standard inflationary models, thus, if the quartic inflationary piece of the Peebles-Vilenkin model is turned into a quadratic one, then the  theoretical values of the parameters $n_s, r$ may enter at $2\sigma$ CL of the plane ($n_s,$ $r$) as reported by \cite{Planck}.

Thus, following this, in the present work we study a quintessential inflationary model after an improvement in the inflationary piece of the well known Peebles-Vilenkin potential. We find that the model 
provides with the theoretical values of the spectral indices in  
agreement with the current observational data about the early universe, and  where the reheating happens due to  the production of heavy massive particles.  These created particles, after their decay into lighter ones and thermalization, form a thermal relativistic plasma whose energy density eventually dominates the one of the inflaton field.
Due to the fact that the model contains a sudden transition from inflation to kination, and since the particles are very massive, this allows us to use the WKB method to calculate analytically the energy density density of the produced particles, and  consequently, assuming a Yukawa-type decay, the corresponding reheating temperature can also be analytically found.
Further, we have also analyzed  its viability in two different situations: when the decay occurs before  and after the end of kination  as well. In both cases, the  maximum reheating temperature lies in the TeV regime. However, if one takes into account the overproduction of Gravitational Waves, following the usual calculation, in order that they do not destabilize the BBN, 
the only viable case is when the decay is before the equilibrium, leading to a reheating temperature lower than $15$ TeV.

We then constrain the quintessence piece of this potential in presence of the latest astronomical datasets aiming to constrain the derived parameters for three different choices of $M$, namely, $M =1 $ GeV, $M = 20$ GeV and $M$ to be a free parameter. Our analyses show that for fixed $M$, the results do not change at all (see Table \ref{tab:resultsI} and Table \ref{tab:resultsII}). In addition to that, the quintessence model behaves in  a similar fashion to other quintessence models (see \cite{Yang:2018xah}, and the references therein), where one can notice the correlations between $H_0$ and $\Omega_{m0}$ while the other set of parameters, such as $(\Omega_{m0}, \sigma_8)$, ($\sigma_8$, $H_0$) are uncorrelated (see Fig. \ref{fig-model1}, Fig. \ref{fig-model2} and Fig. \ref{fig-model3}). However, when we left $M$ to vary in a big interval, this parameter is not constrained at all, at least with the current observational datasets we use in this work. Moreover, we also notice that considering three different scenarios chacaterized by the parameter $M$, the quintessence picce of the model is unable to release the tension on $H_0$ that has appeared from its global \cite{Ade:2015xua} and local measurements \cite{Riess:2016jrr,R18}. A more concrete visualization can be found from Fig. \ref{fig-R16-R18} that establishes such a claim extracted from the present model.

Thus, in summary, we conclude with the comment that the proposed improved version of the well known Peebles-Vilenkin potential needs further examinations with the upcoming observational missions with an aim to investigate the natural consequences of this improvement that are closely related to other issues of the universe evolution as well.

\section*{ACKNOWLEDGMENTS}
The authors thank the referee for some important suggestions and comments to improve the work. This investigation of JH has been supported by MINECO (Spain) grants  MTM2014-52402-C3-1-P and MTM2017-84214-C2-1-P, and  also in part by the Catalan Government 2017-SGR-247. The investigation of WY has been supported by the National
Natural Science Foundation of China under Grants No.  11705079 and No.  11647153.

\section*{Appendix A: One loop energy density for massless particles coupled with gravity}

The vacuum expectation value of the energy density for a  $\chi$-quantum field  coupled to gravity is given by \cite{Bunch}
\begin{eqnarray}\label{energy}
\rho_{\chi}=\frac{1}{4\pi^2 a^4}\int_0^{\infty}\left\{ (|\bar{u}'_k|^2+\omega_k^2(\tau)|\bar{u}_k|^2)+(6\xi-1)\left[ {\mathcal H}( |\bar{u}_k|^2)' -{\mathcal H}^2|\bar{u}_k|^2  \right]   \right\}k^2dk,
\end{eqnarray}
where  $\xi$ is the coupling constant and the prime stands for the differentiation with respect to the conformal time $\tau$, already used in the main text.
The  corresponding Klein-Gordon equation is given by
\begin{eqnarray}
\bar{u}''_k+\Omega_k^2(\tau) \bar{u}_k=0,
\end{eqnarray}
where  we have introduced the notation $\Omega_k^2(\tau)\equiv \omega_k^2(\tau)+(6\xi-1)\frac{a''}{a}$. 
First of all, note that at late times,  the scale factor becomes constant, namely, $a_{\infty}$, and the energy energy density, at late times becomes,
\begin{eqnarray}
\rho_{\chi, \infty}=\frac{1}{4\pi^2 a_{\infty}^4}\int_0^{\infty}(|\bar{u}'_k|^2+\omega_{k,\infty}^2|\bar{u}_k|^2)k^2dk,
\end{eqnarray}
where $\omega_{k,\infty}^2=k^2+m^2 a_{\infty}^2$,  and the dynamical equation is $\bar{u}'_k+\omega_{k,\infty}^2 \bar{u}_k=0$, whose mode solution is 
$\bar{u}_{k,\infty}=\frac{1}{\sqrt{2k}}e^{-i\left(\omega_{k,\infty}\right)\eta} $. Then,
an initial mode, namely, $\bar{u}_{k,0}$, which at late time  becomes $\alpha_k \bar{u}_{k,\infty}+\beta_{k} \bar{u}_{k,\infty}^*$, leading to the following energy density
 \begin{eqnarray}
\rho_{\chi, \infty}=\frac{1}{2\pi^2 a_{\infty}^4}\int_0^{\infty}k\omega_{k,\infty}^2\left(|\beta_k|^2+\frac{1}{2}\right)dk.
\end{eqnarray} 
 The divergent quantity $ \frac{1}{4\pi^2 a_{\infty}^4}\int_0^{\infty}k\omega_{k,\infty}^2 dk  $ corresponds to the Minkowskian vacuum and has to be removed, obtaining finally 
\begin{eqnarray}
\rho_{\chi, \infty}=\frac{1}{2\pi^2 a_{\infty}^4}\int_0^{\infty}k\omega_{k,\infty}^2|\beta_k|^2dk,
\end{eqnarray}
which in the massless case becomes $\rho_{\chi, \infty}=\frac{1}{2\pi^2 a_{\infty}^4}\int_0^{\infty}k^3|\beta_k|^2dk$.

On the other hand, dealing with the quintessential inflation, for the production of massless particles after a phase transition from inflation to a regime with a constant Equation of State (EoS) parameter, such as kination or radiation, the energy density of the created particles can be given by \cite{Spokoiny, pv}: 
 $\rho(\tau)=\frac{1}{2\pi^2 a^4(\tau)}\int_0^{\infty}k^3|\beta_k|^2dk$, which, as we have showed, corresponds to the energy density at very late times, 
 but not to the energy density immediately after the inflation whose correct expression is given by eqn. (\ref{energy}).

To perform the calculations we need the vacuum modes for a linear  EoS ($P=w\rho$) with constant EoS parameter, which are given by  
\begin{eqnarray}
\bar{u}_{k,w}(\tau)=\left\{\begin{array}{ccc}
\sqrt{\frac{\pi\tau}{4}}H^{(2)}_{\nu_w}(k\tau)& \mbox{for} & w>-\frac{1}{3}\\
&  & \\
\sqrt{\frac{-\pi\tau}{4}}H^{(1)}_{\nu_w}(-k\tau)&  \mbox{for} &               w<-\frac{1}{3}, \end{array}\right.
\end{eqnarray}
where    $H^{(1)}_\nu(z)$  and $H^{(2)}_\nu(z)$ are the well-known Hankel functions (see the Chapter $9$ of \cite{abramowitz}) and $\nu_w=\sqrt{\frac{1}{4}+\frac{2(1-3w)}{(1+3w)^2}(1-6\xi)}$ \cite{he}.\\

Since the energy density is a divergent quantity, and in contrary to the massive case it does not seem easy to obtain analytically a well defined quantity, however,  
we can calculate the energy density for massless particles when there is a phase transition form de Sitter to a regime with a constant EoS parameter, only for modes that leave the Hubble radius before
 the phase transition, removing the ultraviolet divergences. In fact, this is the way used in \cite{Damour} to calculate the energy density of the produced particles, and the justification would come from the fact that the modes inside the Hubble radius do not feel gravity, so the modes that are inside the Hubble radius after the phase transition did not feel it and they are not excited enough to produce particles.

For this model,  the conformal Hubble parameter evolves as
 \begin{eqnarray}
 {\mathcal H}=\left\{ \begin{array}{ccc}
 -\frac{1}{\tau}&   \mbox{for} &            \tau<\tau_{kin}<0\\
 &   &\\
 \frac{2}{(1+3w)(\tau-\bar\tau)}&   \mbox{for} &            \tau\geq \tau_{kin},
 \end{array}\right.
 \end{eqnarray}
 where $\bar\tau=\frac{3(1+w)}{(1+3w)}\tau_{kin}$.
 Then,  the vacuum mode before the phase transition is $\bar{u}_{k,-1}(\tau)$ and after it, it becomes $\alpha_k \bar{u}_{k,w}(\tau-\bar\tau)+\beta_k \bar{u}^*_{k,w}(\tau-\bar\tau)$. Matching both modes at $\tau_{kin}$, one obtains
 \begin{eqnarray}
 \alpha_k=
 -iW(\bar{u}_{k,-1}; \bar{u}^*_{k,w}), \quad  
 \beta_k
  =i W(\bar{u}_{k,-1}; \bar{u}_{k,w} ), 
  \end{eqnarray}
 where  $W(f;g)=fg'-f'g$, denotes the Wronskian.
 
In order to simplify the calculations, we consider a transition from de Sitter to kination in the minimally coupled case. During kination, the scale factor behaves as $a=a_{kin}\sqrt{\frac{\tau-\bar\tau}{\tau_{kin}-\bar\tau}}=  a_{kin}\sqrt{{2}{\mathcal H}_{kin}(\tau-\bar\tau)}$. Moreover, in this case, the energy density is given by

\begin{eqnarray}
\rho_{\chi}=\frac{1}{4\pi^2 a^2}\int_0^{\infty}\left(\left|\left(\frac{\bar{u}_k}{a}\right)'\right|^2+k^2\left|\frac{\bar{u}_k}{a}\right|^2\right)k^2dk.
\end{eqnarray} 
In the kination regime, assuming the minimal case, we have $w=0$, then 
for $|z|\ll 1$, a simple calculation leads to
 \begin{eqnarray}
 H^{(1)}_0(z)=  H^{(2)^*}_0(z) \cong 1+\frac{2i}{\pi}\left(\gamma+\ln\left(\frac{z}{2}\right)\right),
 \end{eqnarray} 
 where $\gamma$ is the Euler-Mascheroni constant. In the long wave-length approximation, the vacuum mode is
 \begin{eqnarray}
 \bar{u}_{k,1}(\tau)=\sqrt{\frac{\pi(\tau-\bar\tau)}{4}}\left(
 1-\frac{2i}{\pi}\left(\gamma+\ln\left(\frac{k(\tau-\bar\tau)}{2}\right)\right) \right)
 \nonumber\\=\sqrt{\frac{\pi}{8{\mathcal H}_{kin}}}\left(
 1-\frac{2i}{\pi}\left(\gamma+\ln\left(\frac{k}{4{\mathcal H}}\right)\right) \right)\frac{a(\tau)}{a_{kin}}.
  \end{eqnarray}

 On the other hand, since $\nu_1=\frac{3}{2}$,  before the phase transition one has $u_{k,-1}(\tau)=\frac{e^{-ik\tau}}{\sqrt{2k}}\left(1-\frac{i}{k\tau}\right)$, and thus, the
 Bogoliubov coefficients will be
 \begin{eqnarray}
 \alpha_k=\frac{ie^{-ik\tau_{kin}}}{\sqrt{\pi}}\Bigg[\left(\frac{{\mathcal H}_{kin}}{k}\right)^{3/2}+\frac{1}{2}\left(\frac{{\mathcal H}_{kin}}{k}\right)^{-1/2}{ \left(\gamma+\ln\left(\frac{k}{4{\mathcal H}_{kin}} \right) \right)} \nonumber\\-i
 \left(  \left(\frac{{\mathcal H}_{kin}}{k}\right)^{1/2}+\frac{\pi}{4} \left(\frac{{\mathcal H}_{kin}}{k}\right)^{-1/2} \right)
   \Bigg]
 \end{eqnarray}
 
 \begin{eqnarray}
 \beta_k=\frac{ie^{-ik\tau_{kin}}}{\sqrt{\pi}}\Bigg[\left(\frac{{\mathcal H}_{kin}}{k}\right)^{3/2}+\frac{1}{2}\left(\frac{{\mathcal H}_{kin}}{k}\right)^{-1/2}{ \left(\gamma+\ln\left(\frac{k}{4{\mathcal H}_{kin}} \right) \right)}\nonumber\\-i
 \left(  \left(\frac{{\mathcal H}_{kin}}{k}\right)^{1/2}-\frac{\pi}{4} \left(\frac{{\mathcal H}_{kin}}{k}\right)^{-1/2} \right)
   \Bigg].
 \end{eqnarray} 
Integrating  in the domain $0\leq k\leq {\mathcal H}_{kin}$,  i.e., for modes that leaves the Hubble radius before the phase transition because as we have argued the others do not feel it,  one obtains the following convergent quantity
 \begin{eqnarray}
\rho_{\chi}^{con}(t)=\frac{H_{kin}^4}{32\pi^2}\left(\frac{a_{kin}}{a(t)}\right)^2\left[\left(\frac{a_{kin}}{a(t)}\right)^4+9 -  8\ln\left(\frac{a(t)}{a_{kin}}\right)+\frac{8}{3}\ln^2\left(\frac{a(t)}{a_{kin}} \right)
  \right],
\end{eqnarray} 
which has a completely different behavior from the convergent  quantity normally used to calculate the energy density of the produced particles evolving as \cite{Damour, ford}
\begin{eqnarray}
\frac{1}{2\pi^2 a^4(t)}\int_0^{{\mathcal H}_{kin}}k^3|\beta_k|^2dk\cong \frac{H_{kin}^4}{2\pi^3}\left( \frac{a_{kin}}{a(t)}\right)^4\cong 
10^{-2} H_{kin}^4\left( \frac{a_{kin}}{a(t)}\right)^4.
\end{eqnarray}
 This entails a serious doubt about the way usually used to calculate the vacuum energy density of massless fields, and in particular, the one to calculate the total energy density of  the GWs. Finally, we want to stress that in \cite{glavan}, for massless minimally coupled fields, the authors calculate  the energy density of the produced particles due to a phase transition to radiation using dimensional renormalization, and obtain a result which does not agree (see formula  (83) of \cite{glavan}), for times immediately after the phase transition, with the usual calculation performed in \cite{Damour}. In fact, the agreement between both formulas  is only obtained at very late times,
which seems to indicate that the energy density of the GWs calculated  using the formula  $10^{-2} H_{kin}^4(a_{kin}/a(t))^4$,  could not be used  in eqn.  (\ref{bbnconstraint}) given in Section \ref{subsec-gw2}.

\end{document}